\documentclass[preprint,showpacs,preprintnumbers,amsmath,amssymb,floatfix]{revtex4}

\usepackage{graphicx}
\usepackage{dcolumn}
\usepackage{bm}

\newcommand{\Vp}{{\bm p}}

\makeatletter
\def\lsim{\mathrel{\mathpalette\gl@align<}}
\def\gsim{\mathrel{\mathpalette\gl@align>}}
\def\gl@align#1#2{\lower.6ex\vbox{\baselineskip\z@skip\lineskip.3ex
     \ialign{$\m@th#1\hfill##\hfil$\crcr#2\crcr\sim\crcr}}}
\makeatother

\begin{document}

\preprint{MAP-334}


\title{Lattice calculation of thermal properties
of low-density neutron matter with pionless $NN$ effective field 
theory}

\author{T. Abe${}^{1}$
\footnote{Current address: Center for Nuclear Study, Graduate
School of Science, the University of Tokyo, RIKEN campus, Wako,
Saitama 351-0198, Japan.} and R. Seki${}^{2}$}

\affiliation{
${}^1$
Department of Physics, Tokyo Institute of Technology,
Megro, Tokyo 152-8551, Japan\\
${}^2$ Department of Physics and Astronomy, California State University,
Northridge, Northridge, CA 91330, USA
}

\date{\today}

\begin{abstract}

Thermal properties of low-density neutron matter are investigated
by determinantal quantum Monte Carlo lattice calculations on 3+1
dimensional cubic lattices. Nuclear effective field theory (EFT)
is applied using the pionless single- and two-parameter
neutron-neutron interactions, determined from the $^1S_0$
scattering length and effective range. The determination of the
interactions and the calculations of neutron matter are carried
out consistently by applying EFT power counting rules. The
thermodynamic limit is taken by the method of finite-size scaling,
and the continuum limit is examined in the vanishing lattice
filling limit. The $^1S_0$ pairing gap at $T \approx 0$ is
computed directly from the off-diagonal long-range order of the
spin pair-pair correlation function and is found to be
approximately 30\% smaller than BCS calculations with the
conventional nucleon-nucleon potentials. The critical temperature
$T_c$ of the normal-to-superfluid phase transition and the pairing
temperature scale $T^\ast$ are determined, and the
temperature-density phase diagram is constructed. The physics of
low-density neutron matter is clearly identified as being a
BCS-Bose-Einstein condensation crossover.

\end{abstract}
\maketitle

\section{Introduction}
\label{Introduction}

Neutron matter is of great interest in nuclear physics as a
quantum many-body system. The $^1S_0$ nucleon-nucleon ($NN$)
interaction is strongly attractive, dominating the physics of
neutron matter. The interaction yields the negative (in our
convention) scattering length $a_0$ of an unnaturally large
magnitude ($\approx 20$ fm), with the effective range $r_0$ of a
moderate (natural) size of about twice the pion wavelength ($
\approx 2.8$ fm). The value of $a_0$ implies that the strongly
attractive interaction nearly forms a bound state. By this
pairing, neutron matter is a strongly interacting many-body
system, which must be treated nonperturbatively
\cite{Dean:2002zx}.

The strong neutron pairing generates a pairing gap that creates
superfluidity in neutron matter.  Superfluidity in neutron matter
is of astronomical interest because of the close relation to
the internal structure and thermal evolution of neutron stars
\cite{Heiselberg:2000dn, Lombardo:2000ec}. $^1S_0$ and
${}^3P_2$-${}^3F_2$ superfluidity are believed to be realized in the
inner crust and in the core region of neutron stars, respectively,
and to contribute to the thermodynamic and dynamic properties
of the stars.

Neutron pairing is also considered important for
understanding the structure of neutron-rich unstable nuclei.
Neutron-neutron correlations are expected to be a crucial
ingredient in the weakly bound, surface structure near the
neutron drip line; and for the surface structure, neutron
pairing in neutron matter must be well understood
\cite{Dean:2002zx,Bertsch:1991, Matsuo:2004pr, Hagino:2005we}.

Investigations over many years have provided much understanding of
the physics of thermodynamic properties of neutron matter
\cite{coon,Wiringa:1988tp, Akmal:1998cf}, but reliable
quantitative information of the thermal properties has not
been fully available \cite{Dean:2002zx, Heiselberg:2000dn}. For
example, the $^1S_0$ pairing gap at zero temperature $\Delta$ had
been firmly established {\it in the BCS approximation}, as evident
in the fact that various conventional $NN$ potentials have
provided nearly the same $\Delta_{\rm BCS}$ as a function of
neutron matter density \cite{Elgaroy-1998,schwenk-2007}.
Many-body calculations beyond the BCS mean-field approximation,
however, have yielded $\Delta$ of various magnitudes, generally
smaller than the BCS values, some even by a factor of $2$ or
more. Quantum Monte Carlo calculations, based on a
nonperturbative approach, have also been used on the $\Delta$
determination. The Green's function Monte Carlo (GFMC) method,
quite successful in treating the ground-state properties of finite
nuclei by the use of the conventional $NN$ potentials \cite{GFMC},
has yielded $\Delta$ in the low-density region ($k_F \alt$
0.6 fm${}^{-1}$), smaller than $\Delta_{\rm BCS}$
\cite{Carlson:2007,Gezerlis:2007} but not as small as those
obtained by some of the many-body calculations. Another method
closely related to GFMC, the auxiliary field diffusion Monte Carlo
(AFDMC) method, which is also applied to finite nuclei
\cite{AFDMC-nuclei}, has given $\Delta$ quite close to
$\Delta_{\rm BCS}$ \cite{Fabrocini:2005, Gandolfi:2008} and
significantly larger than the GFMC $\Delta$. We present a more
detailed comparison of these works, including ours, in
Sec.~VII B.

In this paper, we report a quantum Monte Carlo calculation of
$\Delta$ and thermal properties of neutron matter using a method
different from the GFMC and the AFDMC methods.  The difference is
that ours is based on the standard finite-temperature, grand
canonical formulation, while the GFMC and AFDMC methods are
based on essentially zero-temperature formulations, performed for
the ground or specific excited states with a pre-fixed neutron
number. Our calculation may be viewed, in a sense, as a
nonrelativistic hadronic version of lattice QCD calculations, but
it involves different aspects such as those associated with the
large numbers of fermions on the lattice \cite{Lee:2008fa}.
We use a Hamiltonian formulation different from the Lagrangian formulation
commonly used in the lattice QCD calculations.  Our formulation is
not new, as it has been applied in condensed matter physics for 
many years \cite{LohGubernatis1992,dosSantos2003} and has been
also applied in nuclear physics \cite{Muller:1999cp}. This work is
an extension of the latter.

We also use a new ingredient, the $NN$ interaction based on
effective field theory (EFT) \cite{Seki:1998qw, Bedaque:2000kn},
in place of the conventional $NN$ potentials. It is desirable
to include pions \cite{DLee} in the EFT interaction as dynamical
degrees of freedom, representing chiral symmetry and its breaking.
Our objective is twofold: (1) to apply the $NN$ EFT interaction
to the many-nucleon system of neutron matter by properly
applying EFT counting rules, and (2) to determine reliably
the thermal properties of neutron matter and their key
quantities, such as $\Delta$. In the first attempt for
achieving this objective, we have chosen a pionless $NN$ EFT
potential with two parameters. The major consequence of this
choice is that application of our calculation is limited to the
low-density region, $k_F \alt 0.6$ fm$^{-1}$. Even with this
potential, our work has become a relatively large-scale
computation, especially because we take the thermodynamic limit
and examine the continuum limit.  Note that field theoretical
aspects of the general approach of this work were discussed
a few years ago \cite{ck}.

Because the pairing in neutron matter is strong, neutron matter
should be treated as a strongly correlated fermionic system in the
state of BCS-Bose-Einstein condensation (BEC) crossover, which has
been receiving much attention in recent years \cite{crossover}.
Traditionally the pairing in neutron matter has been discussed in
the framework of the BCS approximation \cite{abook}, but the
pairing is too strong for a BCS treatment.  The pairing strength
is characterized by $1/(k_F a_0)$ and corresponds to the BCS
limit with $1/(k_F a_0) \rightarrow -\infty$ and to the BEC limit
with $1/(k_F a_0) \rightarrow + \infty$ \cite{Randeria}. The range
of $1/(k_F a_0)$ in the low-density region investigated in this
work is well in the middle of the two limits, $-0.8 \alt 1/(k_F
a_0) \alt -0.1$, and the magnitude of $1/(k_F a_0)$ becomes
smaller for a higher density. We elaborate on the issue of
crossover in Sec.~VII A.

The limit $1/(k_F a_0) \rightarrow 0$ corresponds to the unitary
limit, to which much attention has been paid lately in the fields
of atomic and condensed-matter physics. A fermion pair in the
unitary limit forms a zero-energy bound state, thereby
yielding a scattering length infinitely long, associated with no
classical scale and expected to have a universal feature. Our
single-parameter EFT description of low-density neutron matter is
close to the unitary limit (rather than to the BCS limit), and we
will discuss the relation between the two in an accompanying
paper \cite{AS-uni}. We emphasize, however, that the close
similarity of the two is restricted to the low-density region of
neutron matter ($k_F \alt 0.3$ fm${}^{-1}$), because additional
EFT parameters and the pionic contributions needed for the
description of the denser region introduce new length scales and
make the physics more complicated than that of the unitary limit.

The outline of this paper is as follows. After the
Introduction of Sec.~I, the basic setup of our calculation is
described in Sec.~II.  In Sec.~III, we present how we
determine the physical quantities of interest in this work, and in
Sec.~IV, we show how we carry out their numerical calculation
by taking the thermodynamic and continuum limits. In Sec.~V,
we discuss how the single- and two-parameter calculations are
matched.  The summary results are shown in Sec.~VI, and
discussions of the key points in this work are given in Sec.~VII. 
A summary of our work is found in Sec.~VIII.  We
include, in Appendix A, a relevant, short discussion on how the
two $NN$ potential parameters are determined by satisfying EFT
counting rules; in Appendix B, a comparison of the physical
sizes of a neutron (Cooper) pair and the computational lattices;
and, in Appendix C, somewhat detailed technical aspects of
our Monte Carlo calculation.

\section{Basic Setup}

\subsection{$NN$ EFT Hamiltonian}

The nuclear EFT Lagrangian is constructed by including all
possible terms allowed by symmetries of the underlying theory of
QCD \cite{weinberg}. The $NN$ potential
from the EFT Lagrangian is written in the momentum expansion form
\begin{equation}
V({\Vp'},{\Vp}) = c_0(\Lambda) +c_2(\Lambda)({\Vp}^2+{\Vp'}^2)+
\cdots - 2c_2(\Lambda){\Vp}\cdot{\Vp'}+ \cdots, \label{EFTpot}
\end{equation}
where ${\Vp}$ and ${\Vp'}$ are the $NN$ center-of-mass
momenta, and $\Lambda$ is the regularization scale.   The terms
not explicitly shown in Eq.~(\ref{EFTpot}) include those in which
pions are treated as a dynamical degree of freedom \cite{pion}.
For the momentum below the pion mass scale, we may neglect the
explicit dynamics of chiral symmetry and its breaking by
truncating Eq.~(\ref{EFTpot}) and including in $c_0$ and
$c_2$ the consequences of the dynamics. In this work, we use this
pionless $S$-wave $NN$ potential with the first two terms in
Eq.~(\ref{EFTpot}). Generally an EFT potential is regarded as an
expansion in terms of $\Vp/\mathcal{Q}$  and $\Vp'/\mathcal{Q}$
with $\mathcal{Q}$ setting the momentum scale of the expansion. In
our pionless potential, we have $\mathcal{Q} \agt m_\pi$ ($m_\pi$,
the pion mass). Note that the potential consists of the
central and spin-dependent parts, as
$c_{c}+\sigma\cdot\sigma^\prime c_\sigma$, with
$\sigma\cdot\sigma^\prime = -3$ for the ${}^1S{}_0$ state (and
$=+1$ for the ${}^3S{}_1$ state, not considered in this work).
We also neglect in this work the $P$-wave interaction term
starting with the ${\Vp}\cdot{\Vp'}$ and the relativistic effects
appearing in $\mathcal{O}(p^4/M^4)$ \cite{bira}.

Regularization is required for the application of
Eq.~(\ref{EFTpot}). On a cubic lattice, the lattice spacing $a$
serves as the regularization scale $\Lambda$, approximately as
\begin{equation}
  \Lambda \sim \frac{\pi}{a}.
\label{lambda1}
\end{equation}
$\Lambda$ should generally be set large, at least larger
than the momentum $p$,
\begin{equation}
  \Lambda > p,
\label{lambda2}
\end{equation}
or better set
\begin{equation}
  \Lambda \agt \mathcal{Q},
\label{lambda3}
\end{equation}
corresponding to $a \alt 4.5$ fm for $\mathcal{Q} \sim m_\pi$
\cite{bira,lepage}. When the two-nucleon interaction is applied to
a many-nucleon system of finite density, an additional constraint
is imposed on the value of $a$, as discussed in Sec.~II C.

On the lattice, the Hamiltonian for our potential takes the
discretized form \cite{ask}
\begin{eqnarray}
  {\hat H} &=& - t \sum_{\langle i,j \rangle \sigma}
  {\hat c}_{i \sigma}^\dagger {\hat c}_{j \sigma}
      + 6t\sum_{i\sigma}
      {\hat c}_{i\sigma}^\dagger {\hat c}_{i\sigma}\nonumber\\
  &+& \frac{1}{a^3}\left[c_0(a)-\frac{6}{a^2}c_2(a)\right] \sum_{i}
  {\hat c}_{i \uparrow}^\dagger
  {\hat c}_{i \downarrow}^\dagger
  {\hat c}_{i \downarrow}
  {\hat c}_{i \uparrow}
  + \frac{1}{2a^5}c_2(a) \sum_{\langle i,j \rangle \sigma \sigma^\prime}
  {\hat c}_{i \sigma}^\dagger {\hat c}_{i \sigma}
  {\hat c}_{j \sigma^\prime}^\dagger
  {\hat c}_{j \sigma^\prime},
  \label{exthubbard}
\end{eqnarray}
where $t = 1/(2M a^2)$, the hopping parameter ($M$ is the
neutron mass), and $\langle i,j \rangle$ denotes a restriction on
the sum to all neighboring pairs. ${\hat c}_{i \sigma}^\dagger$
and ${\hat c}_{i \sigma}$ are the creation and annihilation
operators of the neutron, with $\sigma = \uparrow,
\downarrow$, respectively, at the $i$th site.

The neutron-neutron interaction parameters, $c_0(a)$ and $c_2(a)$,
are determined from the neutron-neutron scattering phase shift,
using the ${}^1S_0$ effective range expansion (ERE),
\begin{equation}
p \cot \delta_0(p) = -\frac{1}{a_0} + \frac{1}{2}r_0 p^2 -P r_0^3
p^4 + \mathcal{O}(p^6), \label{effexp}
\end{equation}
where $P$ is the shape parameter.  By dividing both sides by
$\mathcal{Q}$, we find Eq.~(\ref{effexp}) is an expansion in terms
of the dimensionless quantity $p^2/\mathcal{Q}^2$.  For
$\mathcal{Q} \approx m_\pi$, the coefficients of the expansion
$r_0\mathcal{Q}/2$ and $P(r_0\mathcal{Q})^3$ are of the
natural size $\mathcal{O}(1)$, while the first coefficient is
unnaturally small, $|1/a_0\mathcal{Q}| \ll 1$. Phenomenologically
the sum of the first two terms in Eq.~(\ref{effexp}) agrees well
with the phase shift up to the center-of-mass momentum of
nearly the pion mass $m_\pi \approx 0.7 \;\;{\rm fm}^{-1}$, or
about 40 MeV of the laboratory kinetic energy \cite{bcp} (see also
Ref. \cite{m}). This assures us that $c_0(a)$ and $c_2(a)$ are
safely determined from $a_0$ and $r_0$ for a chosen value of $a$
\cite{sv}.

These interaction parameters are determined by consistently
applying EFT power counting rules in a way different from a mere
phenomenological fitting, as briefly discussed in Appendix A.
Because this determination is one of the crucial steps in this
work, let us note its key point here: {\it $c_2(a)$ and the
contributions of the same order must be treated perturbatively} by
neglecting the $\mathcal{O}([c_2(a)]^2)$ contributions, so that
$\mathcal{O}(p^4/\mathcal{Q}^4)$ contributions are consistently
neglected. Furthermore, {\it to be consistent, $c_2(a)$ and the
contributions of the same order must also be treated
perturbatively in the neutron matter calculations}.  In the next
subsection, we discuss how this treatment is formulated for the
neutron matter calculation.

In this work, we carry out the neutron matter calculation using
Eq.~(\ref{exthubbard}) in two different ways: the leading-order 
(LO) calculation, in which the $c_2(a)$ contribution and the
contributions of the same order are neglected, and the
next-to-leading-order (NLO) calculation, in which they are
included. The LO and NLO calculations are expected to yield
somewhat different physics, because Eq.~(\ref{exthubbard}) is the
Hamiltonian of the attractive Hubbard model for the LO
calculation, and it is the Hamiltonian of an extended
attractive Hubbard model for the NLO calculation \cite{ask}. 
With the neglect of $\mathcal{O}(p^2/\mathcal{Q}^2)$, the LO
calculation involves the neutrons of low momenta and should
be applicable to a low-density region of neutron matter without
the perturbative treatment.

An important issue in this work is the density at which the
LO and NLO results should be matched.  The ERE of
Eq.~(\ref{effexp}) suggests that the center-of-mass momentum of an
interacting neutron {pair is less} than $\sqrt{2/(|a_0| r_0)}
\approx 0.20 \;{\rm fm}^{-1}$ at the matching density.  As a
rough estimate, it may be feasible to identify the Fermi momentum
$k_F$ as this momentum and to estimate the density from it
\cite{Dean:2002zx}, but for a rigorous matching, the LO and NLO
neutron matter calculations should be carried out for some common
densities and their results compared. As it is desirable to
avoid excess computer time, we use in this work the following
procedure: we carry out the LO and NLO calculations at the common
density of $k_F =$ 0.3041 ${\rm fm}^{-1}$, where we expect
the two results will certainly differ, and then perform similar
calculations by lowering the density so as to identify the density
that yields the same LO and NLO results (within the statistical
uncertainties). The matching using this procedure is
elaborated in Sec.~V.

\subsection{Determinantal quantum Monte Carlo computation}
\label{Subsection:DQMC}

We follow a lattice Hamiltonian formulation, somewhat
different from the Lagrangian formulation usually used in
lattice QCD \cite{LatticeQCD}. Instead of using the representation
in terms of coherent-state Grassmann variables, we use the number
representation, working with the lattice Fock space $\langle n|$
using the creation and annihilation operators of the neutrons.
Our treatment is the same as that used in 
Refs.~\cite{lang,Muller:1999cp,ask} and is commonly used in
condensed-matter physics \cite{LohGubernatis1992,dosSantos2003}
under the determinantal quantum Monte Carlo (DQMC) method.

We carry out neutron matter calculations using the Hamiltonian of
Eq.~(\ref{exthubbard}) in the method of grand canonical ensemble.
The Monte Carlo computation is carried out for various
values of the chemical potential $\mu$, and the $\mu$ dependence
is converted to the density dependence by determining the
densities by the average over ${i,\sigma}$ of $\langle
{\hat c}^\dagger_{i\sigma} {\hat c}_{i\sigma}\rangle$ for various values of
$\mu$.

For many-nucleon systems, the Hamiltonian (\ref{exthubbard})
should also include three-nucleon interactions.  By EFT power
counting rules, the interactions are to be treated generally as
the LO order in the pionless case, and they play a significant role
when a three-nucleon bound state such as the triton can be formed
\cite{bk}. In neutron matter, however, the three-neutron system
has no bound state, and the three-neutron interactions appear at a
higher order because the Fermi statistics prohibit the LO diagram
of three neutron from being at the same spatial point with the
momentum-independent vertex. As the interactions would also affect
the two-neutron pairing indirectly, we expect that the
interactions would affect the observables of our interest
relatively weakly and defer the issue to a future investigation
by neglecting them in this work.

We write the partition function as
\begin{equation}
\mathcal{Z}(T,\mu) \equiv \langle n|{\hat U}(\beta)|n \rangle,
\label{partition}
\end{equation}
where ${\hat U}(\beta)$ is the (imaginary time) evolution
operator, and the trace implied in Eq.~(\ref{partition}) is over
all possible nucleon configurations on the lattice $\langle
n|$. Using the Trotter-Suzuki approximation, we express ${\hat
U}(\beta)$ as
\begin{equation}
{\hat U}(\beta) =\mathcal{T}{\rm
exp}\left[-\sum_{\tau_t=1}^{N_t}\Delta\beta \left({\hat H}
-\mu\sum_{i\sigma}{\hat c}_{i \sigma}^\dagger{\hat c}_{i
\sigma}\right)\right] \equiv \mathcal{T}\Pi_{\tau_t=1}^{N_t} {\hat
U}(\Delta\beta) \label{latticeU}
\end{equation}
by the temporal discretization  $\beta = \Delta\beta N_t$,
with $N_t$ being the number of time slices. In
Eq.~(\ref{latticeU}), ${\hat H}$ is the two-parameter NLO
Hamiltonian of Eq.~(\ref{exthubbard}), and $i$ is actually an integer
vector specifying the location of a site with its component
ranging as $[-aN_s^{1/3}/2,aN_s^{1/3}/2]$. The $\tau_t$
dependence of $\hat{H}$ and ${\hat U}(\Delta\beta)$ is solely
through ${\hat c}^\dagger$ and ${\hat c}$, as seen from
Eq.~(\ref{exthubbard}). The last expression in
Eq.~(\ref{latticeU}) is thus a product of ${\hat U}(\Delta\beta)$
operators, each having the same form and depending on $\tau_t$
implicitly.

To cast $\mathcal{Z}(T,\mu)$ in a form amenable to Monte
Carlo computation of the fermion integration, we express the
two-nucleon interaction of $\hat{H}$ in a single-nucleon
interaction form by applying the Hubbard-Stratonovich
transformation 
\begin{equation}
e^{+A{\hat n}_i^2} = \sqrt{\frac{A}{\pi}} \int d\chi_i
e^{-A(\chi_i^2-2\chi_i{\hat n}_i)} \label{HST}
\end{equation}
for a constant $A$ with Re$(A) > 0$.  Here, $\chi_i$ is an
auxiliary scalar field at the $i$th site, and ${\hat n}_i$ is the
density operator defined as ${\hat n}_i \equiv {\hat n}_{i
\uparrow}+{\hat n}_{i \downarrow}$ ( ${\hat n}_{i \sigma}\equiv
{\hat c}_{i \sigma}^\dagger {\hat c}_{i \sigma}$, the number
operator with the spin $\sigma$ at the $i$th site). $\hat{H}$
is divided into two parts,
\begin{equation}
  {\hat H} \equiv \left[{\hat H}_s
  + \frac{1}{2a^3}c_0^{(0)}(a)\sum_i{\hat n}_i^2\right]
  +{\hat H}^\prime,
  \label{hubbardR}
\end{equation}
where
\begin{eqnarray}
{\hat H}_s&\equiv&- t \sum_{\langle i,j \rangle \sigma}
  {\hat c}_{i \sigma}^\dagger {\hat c}_{j \sigma}+ \left[6t
- \frac{1}{2a^3}c^{(0)}_0(a)\right]\sum_i{\hat n}_i\nonumber\\
{\hat H}^\prime &\equiv& \frac{1}{2a^5}c_2(a)\sum_{\langle
i,j \rangle}{\hat n}_i{\hat n}_j +\frac{1}{2a^3}\left[\Delta
c_0(a) -\frac{6}{a^2}c_2(a)\right]\sum_i\left({\hat
n}_i^2 -{\hat n}_i\right). \label{latticehp}
\end{eqnarray}
Here, $c_0(a)$ is expressed as a sum of the LO part $c_0^{(0)}(a)$
and the NLO part $\Delta c_0(a)$, which are defined in
Eqs.~(\ref{c00}) and (\ref{c02eft}), respectively, with
$\Lambda = \pi/a$.

We introduce $\hat{H}_0(\chi)$, the LO single-nucleon Hamiltonian
interacting with the external scalar field $\chi \equiv
\{\chi_i\}$,
\begin{equation}
{\hat H}_0 (\chi) \equiv {\hat H}_s
+\frac{1}{a^3}c_0^{(0)}(a)\sum_i\chi_i{\hat n}_i. \label{hath0}
\end{equation}
In terms of $\hat{H}_0(\chi)$, ${\hat U}(\Delta\beta)$ is written
as
\begin{eqnarray}
{\hat U}(\Delta\beta) &=&\int d [\chi] {\rm exp}
\left[+\frac{\Delta\beta}{2a^3}c_0^{(0)}(a)\sum_i\chi_i^2\right]
{\rm exp} \left\{-\Delta\beta\left[{\hat
H}_0(\chi)+{\hat H}^\prime-\mu\sum_i{\hat n}_i\right]\right\}\nonumber\\
&\approx& \int d [\chi]
e^{+\frac{\Delta\beta}{2a^3}c_0^{(0)}(a)\sum_i\chi_i^2}
(1-\Delta\beta{\hat H}^\prime)e^{-\Delta\beta\left[{\hat
H}_0(\chi)-\mu\sum_i{\hat n}_i\right]}, \label{delu}
\end{eqnarray}
where the measure is defined as $d [\chi] \equiv d\chi_1
d\chi_2\ldots$ with a constant factor generated by the
Hubbard-Stratonovich transformation.  We emphasize that ${\hat
H}^\prime$ is defined to be of the NLO and is treated
perturbatively in the second step of Eq.~(\ref{delu}).

We thus obtain
\begin{eqnarray}
\mathcal{Z}(T,\mu) &\approx&\int
d[\chi]\mathcal{T}\Pi_{\tau_t=1}^{N_t}
e^{+\frac{\Delta\beta}{2a^3}c_0^{(0)}(a)\sum_i\chi_i^2}\langle n|
(1-\Delta\beta{\hat H}^\prime)e^{-\Delta\beta\left[{\hat
H}_0(\chi)-\mu\sum_i{\hat n}_i\right]}|n \rangle\nonumber\\
&\equiv& \int d[\chi]G(\chi)\langle n|{\hat U}_{\chi}(\beta)|n
\rangle, \label{latpartition}
\end{eqnarray}
where
\begin{eqnarray}
G(\chi) &\equiv& \Pi_{\tau_t=1}^{N_t}e^{+
\frac{\Delta\beta}{2a^3}c_0(a)\sum_i\chi_i^2}\nonumber\\
{\hat U}_{\chi}(\beta) &=& \Pi_{\tau_t=1}^{N_t} {\hat
U}_{\chi}(\Delta\beta) \equiv\Pi_{\tau_t=1}^{N_t}
(1-\Delta\beta{\hat
H}^\prime) e^{-\Delta\beta\left[{\hat
H}_0(\chi)-\mu\sum_i{\hat n}_i\right]}. \label{latg}
\end{eqnarray}
Note that the time-ordering (sequential) integration over $[\chi]$
is understood in the last expression of Eq.~(\ref{latpartition}).
In accordance with Eq.~(\ref{delu}), the factor $(1-\Delta\beta {\hat
H}^\prime)$ in ${\hat U}_{\chi}(\Delta\beta)$ of Eq.~(\ref{latg})
is to be evaluated by the use of the nucleon lattice configuration
resulting from the ${\rm exp}\{-\Delta\beta[{\hat
H}_0(\chi)-\mu\sum_i{\hat n}_i]\}$ operation at $\tau_t$, and that
this procedure is repeated successively from $\tau_t=0$ to $N_t$.
This step is vital in the computation for the perturbative
treatment of ${\hat H}^\prime$.  We make a technically important
note: because of the perturbative treatment of ${\hat
H}^\prime$, the number of the auxiliary fields for the NLO
calculation remains as $N_s N_t$, the same as for the LO
calculation. If ${\hat H}^\prime$ were not treated perturbatively,
$4 \times N_s N_t$ more of $\{\chi_i\}$ would have been needed
owing to the derivative interactions, and the Monte Carlo
computation would have required more time by nearly an order
of magnitude.

The trace of the single-particle evolution operator ${\hat
U}_{\chi}(\beta)$ is expressed in terms of the single-particle
matrix representation of the operator, ${\bm U}_{\chi}(\beta)$, as
\cite{LohGubernatis1992,Muller:1999cp, dosSantos2003,lang}
\begin{equation}
\langle n|{\hat U}_{\chi}(\beta)|n \rangle = {\rm det}\left[1+{\bm
U}_{\chi}(\beta)\right] \equiv \xi(\chi). \label{matrix}
\end{equation}
The expectation value of the (static) operator
$\mathcal{O}({\hat c}^\dagger, {\hat c})$ at $T=1/\beta$ is then
obtained from
\begin{eqnarray}
\langle \mathcal{O}({\hat c}^\dagger, {\hat c})\rangle
&=&\frac{1}{\mathcal{Z}(T,\mu)}\int d[\chi] G(\chi)\langle n|
\mathcal{O}({\hat c}^\dagger, {\hat c}){\hat U}_{\chi}(\beta)|n
\rangle \nonumber\\
&\equiv& \frac{\int d[\chi]G(\chi)\langle\mathcal{O}(\chi)\rangle
\xi(\chi)}{\int d[\chi]G(\chi)\xi(\chi)}, \label{montecarlo}
\end{eqnarray}
where $\langle\mathcal{O}(\chi)\rangle$ is
\begin{equation}
\langle\mathcal{O}(\chi)\rangle \equiv \frac{\langle
n|\mathcal{O}({\hat c}^\dagger, {\hat c}){\hat U}_{\chi}(\beta)|n
\rangle} {\langle n|{\hat U}_{\chi}(\beta)|n \rangle}
\end{equation}
and can be evaluated in terms of ${\bm U}_{\chi}(\beta)$
using Eq.~(\ref{matrix}), as shown in Refs.
\cite{LohGubernatis1992,Muller:1999cp,dosSantos2003,lang}.

Equation~(\ref{montecarlo}) is now amenable to a Monte Carlo
integration by treating $|G(\chi)|$ or $|G(\chi)\xi(\chi)|$ as a
weight.  Our Monte Carlo computation is the same as that
used in Ref.~\cite{Muller:1999cp}, supplemented by a
matrix-decomposition stabilized method for low-temperature
computations \cite{LohGubernatis1992,dosSantos2003}.  

Before closing this subsection, we make a relevant comment.  In
the procedure just described, we reduced the
original Hamiltonian $\hat{H}$ of Eq.~(\ref{exthubbard}) to the
single-nucleon Hamiltonian ${\hat H}_0$ (with ${\hat H}^\prime$)
of Eq.~(\ref{hath0}) in terms of the density operators
$\{{\hat n}_i\}$, as in Eq.~(\ref{hubbardR}). The choice of the
density operators in this step may seem natural, but it is not
required for the reduction to an effective single-nucleon
Hamiltonian because of the arbitrariness in the path integral
formulation.  In fact, we can choose a combination of pairing
operators and density operators, leading to a
Hartree-Fock-Bogoliubov (HFB) type Hamiltonian
\cite{negele,lang}.

\subsection{Lattice spacing toward thermodynamic and continuum limits}

Neutron matter is a strongly correlated fermion system. On a
three-dimensional cubic lattice, the correlation length
resulting from the simulation, $\xi$, satisfies
\begin{equation}
 a \ll \xi \alt L,
\label{xi}
\end{equation}
where $L \equiv a N_s^{1/3}$ is the physical dimension of the
cubic lattice. $\xi$ is the length scale in which the collective
state is realized in the simulation and is different from the
size of a neutron pair (a Cooper pair) in the state
$\xi_{\rm cp}$. Note that, confusingly, $\xi_{\rm cp}$ has often been
referred to terms similar to $\xi$. In Appendix B, we compare the
physical sizes of the neutron pair simulated and the lattice
spaces used.

To obtain a physically meaningful result, we seek for
$\xi$ and for the expectation values of other quantities, in the
continuum limit $a \rightarrow 0$ and in the thermodynamic limit
$L \rightarrow \infty$.  The clear procedure for achieving
both limits is to do the former with $L$ fixed (for obtaining
results insensitive to the lattice structure), and then to do the
latter (using finite volume corrections), as is usually done
in lattice QCD calculations \cite{LatticeQCD}.

In our calculation of the many neutron system, each meaningful
configuration must consist of neutrons fewer than $N_s$, so that
the calculation properly describes the interacting neutron system
{\it in free space}.  This requirement is crucial in general for
the simulation of a system of many fermions, and we find that the
requirement complicates the straightforward approach of
achieving the above two limits.  Note that lattice QCD
calculations have not yet dealt with cases of such high
baryon-density states.

Let us elaborate on this requirement.  Consider setting up a
classical lattice configuration. When $N_f$ neutrons are placed on
a lattice of volume $a^3 N_s$, the neutron density $\rho$ is
\begin{equation}
\rho \equiv \frac{N_f}{N_s a^3} \equiv \frac{n}{a^3},
\label{fraction}
\end{equation}
which defines the lattice-filling fraction (or more descriptively,
the site-occupation fraction), $n \equiv N_f/N_s$.  $n$ denotes
the fraction of the lattice sites occupied by the neutrons. Note
that the complete filling of the lattice occurs with $n=2$ owing
to the spin degree of freedom.  Classically, $n$ can simply be
chosen, while in our quantum-mechanical, grand canonical
calculation, it is determined from $\sum_{i,\sigma} \langle
{\hat c}_{i\sigma}^\dagger {\hat c}_{i\sigma} \rangle$, which is
computed for a fixed value of $a$ and $\mu$.

Mathematically, for a finite nucleon density, Eq.~(\ref{fraction})
implies
\begin{equation}
   n(\mu) \rightarrow 0
\label{nlimit}
\end{equation}
as $a \rightarrow 0$.  Physically, these limits simulate the
free-space environment, because the smaller $n$ is, the more
vacant sites are available, allowing more feasible excitations
to be realized. To determine thermal quantities as
a function of neutron matter density, we consider achieving the
limits to be vital and take the limit of Eq.~(\ref{nlimit})
as the continuum limit. Note that this procedure is similar to,
but different from, the one recently proposed for the unitary
limit problem \cite{burovski}, in that we keep the density $\rho$
finite as we approach the continuum limit, but the $k_F
\rightarrow 0$ limit is taken in Ref. \cite{burovski}.

Once we decide to take Eq.~(\ref{nlimit}) as the continuum limit,
we have to use different values of $a$ for different densities to
satisfy the regularization scale requirement, Eqs.~(\ref{lambda2})
and (\ref{lambda3}), of the EFT.  The procedure becomes
complicated in order to satisfy all these requirements, but at the
same time it has to be durable in practice.  We have decided
to use the following procedure. First, we choose an
appropriate $n$ value that is small enough yet reasonably
durable.  Second, for this $n$, we choose a set of the
representative nucleon densities for the computation and a
set of appropriate $a$ values for them.  We call the set the
standard parameter set, and we list them in Sec.~II D.
Third, after we complete the computation for the standard set, we
perform the computation by varying the lattice size, so as to take
the thermodynamic limit. Fourth, we vary $n$ to examine the
continuum limit as $n \rightarrow 0$.

In the rest of this subsection, we discuss the first step, how we
choose $n$ for the standard set. As an estimate, take the
Fermi gas model. In terms of the Fermi momentum $k_F$, $n$ for
neutron matter is written as
\begin{equation}
    n = (a k_F)^3/(3\pi^2) \rightarrow 0.
\label{nak}
\end{equation}
To keep $n$ independent of $a$ for various densities, we
should have $a \propto 1/k_F$.  Note that the excitation energies
of the neutron matter of interest are about an order of magnitude
less than the Fermi energy, as seen in Sec.~V, and are safely
ignored in this estimate.

The smallness of $n$ is achieved by making $a$ small, or $\Lambda$
large. If we take Eq.~(\ref{lambda2}), Eq.~(\ref{lambda1}) with $p
\sim k_F$ yields
\begin{equation}
   \pi > a k_F.
\label{cutoff1}
\end{equation}
Equations (\ref{nak}) and (\ref{cutoff1}) yield a rather loose
estimate of $n < 1$. We can obtain a more realistic limit from
the observation that the lattice discretization amounts to the
replacement
\begin{equation}
  {\bf p}^2 \rightarrow \frac{2}{a^2} \sum_{i=1}^{3}
  \left[ 1 - \cos (a p_i) \right] = {\bf p}^2 + {\cal O} (a^2 {\bf
  p}^4),
\end{equation}
for example, in the neutron propagator. This observation suggests
that the left-hand side of Eq.~(\ref{cutoff1}) is more like unity
instead of $\pi$, and we obtain the inequality
\begin{equation}
  (3\pi^2)^{-1/3} > n.
\end{equation}
This choice of $n$ does not require a large $\Lambda$ to satisfy
Eq.~(\ref{lambda3}), but it does for Eq.~(\ref{lambda2}).

The preceding consideration leads us to set $n =1/4$ (or 1/8 of
the full filling of the lattice), as a practical compromise.
Other parameters also need to be chosen. In the following
subsection, we discuss how they are chosen and list all parameter
values in the standard parameter set.

\subsection{Standard parameter set}

The standard set of the potential parameters is shown in
Table~\ref{table:Potential_Parameters}.  We choose the set by the
following steps. First, we choose the values of the Fermi momentum
$k_F$, representing the neutron matter density, as shown in the
first and second columns. Second, the values of $a$ are determined
from $(a k_F)^3/(3 \pi^2) = n = 1/4$ and at the same time by
ensuring that the $a$ values provide reasonable EFT regularization
scales. Third, the values of $c_0$ and $c_2$ are determined from
$a_0$ and $r_0$ using the $a$ values in Eqs.~(\ref{c02eft})
and (\ref{c00}) with $\Lambda=\pi/a$. The Monte Carlo
calculations are carried out using the $c_0$ and $c_2$ values by
tuning the chemical potential $\mu$, so that the resultant neutron
matter densities by the Monte Carlo computation are the $\rho$
values listed in the third column in the unit of the normal
nuclear density $\rho_0 = 0.16$ fm$^{-3}$. We emphasize that
these $\rho$ values are expressed in terms of the $k_F$ values in
the first and second column exactly as
\begin{equation}
\rho = k_F^3/(3\pi^2).
\label{realD}
\end{equation}
Throughout this work, we use $k_F$ defined through
Eq.~(\ref{realD}) for specifying the {\it quantum-mechanically
computed density, $\rho$}, of neutron matter as the interacting
fermion system.

\begin{table}[htbp]
\caption[Standard parameter set.] {Standard parameter set.}
\label{table:Potential_Parameters}
    \begin{center}
    \begin{tabular}{ccccccc}
      \hline
      \hline
       &$k_F$ (MeV) & $k_F$ $({\rm fm}^{-1})$  & $\rho$ ($\rho_0$)
      & $a$ (fm) & $c_0/(a^3t)$ & $c_2/(a^5t)$ \\
      \hline
       LO &  \, $15$ & \, $0.07602$ & $9 \times 10^{-5}$ & $25.64$  &  \, $-5.308$ & -- \\
       LO &  \, $30$ & $0.1520$  & $7 \times 10^{-4}$ & $12.82$  & \, $-6.354$ & -- \\
       LO &  \, $60$ & $0.3041$  & $6 \times 10^{-3}$ &  \ \, $6.409$ & \, $-7.049$ & -- \\
      NLO &  \, $60$ & $0.3041$  & $6 \times 10^{-3}$ &  \ \, $6.409$ & \, $-9.646$ & $0.3684$ \\
      NLO &  \, $90$ & $0.4561$  & $2 \times 10^{-2}$ &  \ \, $4.273$ & $-11.074$ & $0.5139$ \\
      NLO & $120$ & $0.6081$  & $5 \times 10^{-2}$ &  \ \, $3.205$ & $-12.343$ & $0.6555$ \\
      \hline
      \hline
    \end{tabular}
  \end{center}
\end{table}

\section{Determination of $\Delta$, $T_c$, and $T^\ast$
from the pairing correlation function}
\label{Numerical_Results_and_Discussions}

In this work, we focus on the determination of three quantities:
the $^1S_0$ pairing gap at $T \approx 0$, $\Delta$; the critical
temperature $T_c$ of the normal-to-superfluid phase
transition; and the pairing temperature scale $T^\ast$.  The
latter two will be used to obtain the density-temperature phase
diagram, and all quantities will be calculated from correlation
functions, the first two from the pair-pair correlation function
and the third from the magnetic susceptibility (the spin-spin
correlation).

\subsection{Pairing gap $\Delta$}
\label{subsec:P_s}

\begin{figure}[htbp]
\begin{center}
\includegraphics[width=100mm]{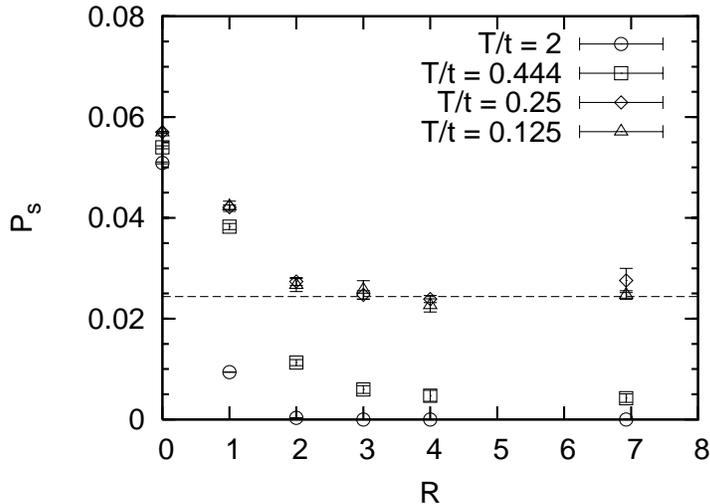}
\caption{Spin pair-pair correlation function $P_s$ as a
function of the lattice separation $R$ for the lattice size $N_s =
8^3$ for the site-occupation fraction $n = 0.25$ at $k_F = 30$
MeV. The DQMC results are shown with statistical uncertainties for
$T/t = 2.0$, $0.444$, $0.25$, and
$0.125$ in the unit of hopping amplitude $t = 0.126$
MeV. The dashed line is the asymptotic value of $P_s = 0.0244(6)$
extracted from the values for $R=$ 4 and $4\sqrt{3}$ at $T/t=$
0.25, 0.125, and 0.0625 (not shown).}
\label{Fig:P_s}
\end{center}
\end{figure}

$\Delta$ is determined directly from the off-diagonal long-range
order (ODLRO) of the spin pair-pair correlation function $P_s$
\cite{Guerrero},
\begin{eqnarray}
  P_s(R) &=& \frac{1}{N_s} \sum_i
  \langle \hat{\Delta}_{i+R}^\dagger \hat{\Delta}_i \rangle \nonumber \\
  &=& \frac{1}{N_s} \sum_{i,\;j=i+R} \left( \delta_{i j} - G_{j i} \right)^2,
\end{eqnarray}
where $\hat{\Delta}_i \equiv {\hat c}_{i\uparrow}{\hat c}_{i\downarrow}$ is the
two-neutron spin-pairing operator at the $i$th site, and $R$ is
the separation of the neutron pairs in the lattice spacing unit
and has no dimension. Note that $G_{ij} \equiv G_{ij}^\sigma =
\langle {\hat c}_{i\sigma} {\hat c}_{j\sigma}^\dagger \rangle$ for
$\sigma = \uparrow$, $\downarrow$ in the attractive Hubbard model.
$P_s(R)$ decays rapidly in $R \approx$ 1 or 2 and takes a
diminishing asymptotic value. When a long-range order exists
between neutron pairs, the asymptotic value is finite, that is,
the signature of the ODLRO. Figure~\ref{Fig:P_s} illustrates this
behavior.

In Fig.~\ref{Fig:P_s}, $P_s(R)$ is calculated for 14 values of
$T/t$ between 2.0 and 0.0625; but for clarity, only the selected
values of $T/t$ are shown. Note that the integer points of 
$R = 1$-$4$ arise from the lattice points in the side of the cubic,
while the largest $R = 4\sqrt{3} \approx 7$ comes from the
midpoint of the diagonal line in the cubic, which has the
displacement vector $\langle 4,4,4 \rangle$. The values for $R
\geq 3$ are found to be quite close to each other at the lowest
three temperatures, $T/t=$ 0.25, 0.125, and 0.0625.  The values at
$R=$ 4 and 7 are averaged, yielding $P_s(T \approx 0, R \gg 1)$.
$\Delta$ is then determined from
\begin{equation}
  \Delta = \frac{|c_0|}{a^3} \sqrt{P_s(T \approx 0, R \gg 1)}.
\label{Eq:Delta_ourdefinition}
\end{equation}
Similar procedures are applied for different $N_s$ and $k_F$.
The errors from the fit hereafter are estimated by a constrained 
least-squares method.

Note that, as seen in Fig.~\ref{Fig:CDelta8_kF30} of the next
subsection, the critical temperature is $T_c/t = 0.335(1)$, and
the behavior of $P_s(R)$ at $T_c$ is similar to $T/t=0.25$ in
Fig.~\ref{Fig:P_s}.  We caution the reader that the $\Delta$
thus determined is not our final value but is the value for $N_s
=8^3$ and $n=1/4$ at $k_F=$ 30 MeV. Using $\Delta$'s for various
values of $N_s$ and $n$ at each $k_F$, we determine $\Delta$ at
the thermodynamic and continuum limits by the further analysis
described in Sec.~IV.  The same caution is applied to the
determination of $T_c$ and $T^\ast$ in the following subsections.

\subsection{Critical temperature $T_c$}
\label{subsec:Determination_Tc}

$T_c$ of the normal-to-superfluid phase transition is determined
from the spin pair-pair correlation sum \cite{Moreo:1991,
dosSantos:1993, dosSantos:1994, Sewer}
\begin{eqnarray}
  C_\Delta (T) &=& \frac{1}{N_s} \sum_{i,j}
  \langle {\hat \Delta}_i {\hat \Delta}_j^\dagger
  + {\hat \Delta}_i^\dagger {\hat \Delta}_j \rangle
  \nonumber \\
  &=& \frac{1}{N_s} \sum_{i,j} \left[
  \left(G_{ij}\right)^2 + \left( \delta_{ij} - G_{ji} \right)^2 \right].
\end{eqnarray}

\begin{figure}[htbp]
\begin{center}
\includegraphics[width=100mm]{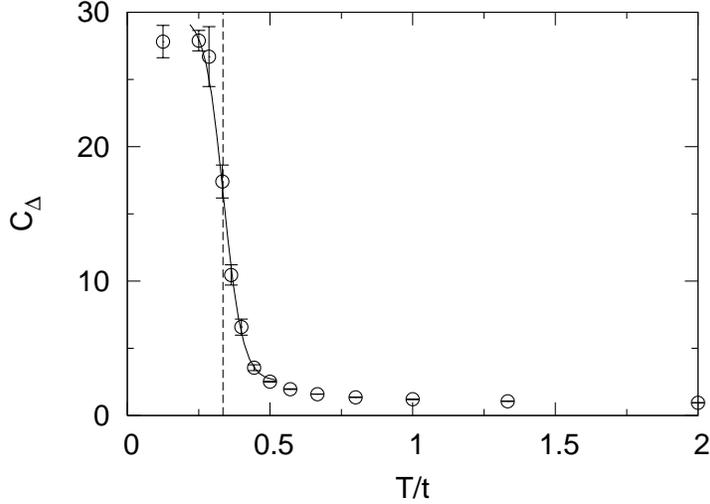}
\caption{Spin pair-pair correlation sum $C_\Delta$ as a
function of temperature $T$ in the unit of the hopping parameter
$t$. The Monte Carlo data with the statistical uncertainties are
shown for the case of $N_s = 8^3$ and $k_F = 30$ MeV. The vertical
dashed line signifies $T_c(N_s = 8^3) = 0.335(1) t$, corresponding
to the inflexion point of the interpolated curve of
$C_\Delta(T/t)$, Eq.~(\ref{tstarfit}) with $C_1 = 13.6 \pm 1.0$,
$C_2 = 15.1 \pm 2.9$, and $C_3 = 16.2 \pm 0.6$. The interpolated
curve is shown as the solid curve.} \label{Fig:CDelta8_kF30}
\end{center}
\end{figure}

$T_c$ is extracted from the inflexion point of $C_\Delta (T)$.
Figure~\ref{Fig:CDelta8_kF30} illustrates a typical case of
$C_\Delta$ as a function of $T/t$, for $k_F = 30$ MeV and $N_s =
8^3$.  In the figure, the inflexion point is at $T_c/t =
0.335(1)$, or $T_c = 0.0423(1)$ with $t = 0.1261$ MeV. The
inflexion point is determined by an interpolation that fits the
Monte Carlo data with an assumed function,
\begin{equation}
C_\Delta (T) = - C_1 \tanh[C_2(T - T_c)/t] + C_3,
\label{tstarfit}
\end{equation}
where $C_1 = C_\Delta(T=0)/2$, $C_2$, and $C_3$ are free constant parameters.

\subsection{Pairing temperature Scale $T^\ast$}
\label{subsec:Determination_Tast}

\begin{figure}[htbp]
\begin{center}
\includegraphics[width=100mm]{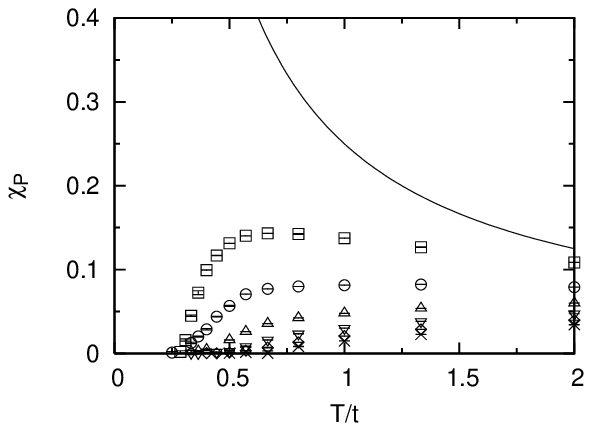}
\caption{Pauli spin susceptibility $\chi_P$ as a function of
temperature $T$ in the unit of hopping parameter $t$ for $N_s=
4^3$. The solid curve is the free fermion gas limit ($|c_0|/(a^3t)
\rightarrow 0$) of $\chi_P(T)$, $\approx n(1-n/2)/T$ (with the
filling fraction $n$) \cite{Sewer:PhDthesis}. In comparison to
this, the cases for $|c_0|/(a^3t) = 2$, $4$, $6$, $8$, $10$, and $12$
are shown in the increasing order of the interaction strength,
from top to bottom.} \label{Fig:chiP4_2}
\end{center}
\end{figure}

As the temperature increases, the long-range order of the
superfluidity disappears at $T_c$.  Above $T_c$, the spin pairing
still remains, however, without generating the long-range order,
and as the temperature increases further, the pairing eventually
disappears. Though the process of the pairing disappearance is
expected to be a continuous process, we may identify the
temperature below which the pairing can be viewed as still strong.
Following a practice in condensed-matter physics \cite{Sewer,
dosSantos:1994}, we denote the temperature as the pairing
temperature scale $T^\ast$ and determine it from the temperature
dependence of the Pauli spin susceptibility $\chi_P$. When the
($S$-wave singlet) spin pairing is weakened, the spectral weight
of low-energy spin excitations is reduced, and the spin response
weakens.  $\chi_P$ is a good quantity for studying this
transition, since the $\chi_P$ of a free fermion gas diverges as
$T \rightarrow 0$, while it vanishes for an interacting fermion
gas, as illustrated in Fig.~\ref{Fig:chiP4_2}.

$\chi_P$ is given by
\begin{eqnarray}
  \chi_P (T,N_s) &=& \frac{1}{T}\frac{1}{N_s} \sum_{i,j}
  \langle \mbox{\boldmath $S$}_i
  \cdot \mbox{\boldmath $S$}_j \rangle
  \nonumber \\
  &=& \frac{1}{T}\frac{1}{N_s} \sum_{i,j}
  2 G_{ij} \left( \delta_{ij} - G_{ji} \right),
\end{eqnarray}
where $\mbox{\boldmath $S$}_i = \sum_{\mu,\nu = \uparrow,
\downarrow} c_{i\mu}^\dagger \mbox{\boldmath $\sigma$}_{\mu\nu}
c_{j\nu}$ and $\mbox{\boldmath $\sigma$}$ is the Pauli vectorial
matrices.  $T^\ast$ is determined by identifying the maximum point
of $\chi_P$ as a function of $T$ \cite{Sewer, dosSantos:1994}, as
discussed in the following.

\begin{figure}[htbp]
\begin{center}
\includegraphics[width=100mm]{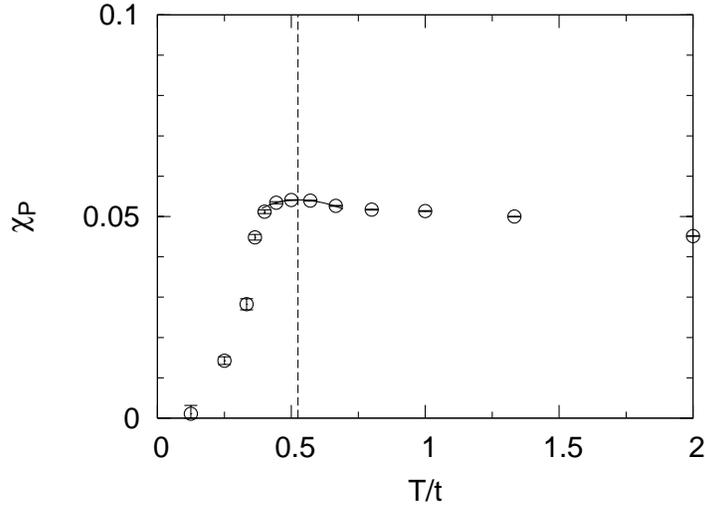}
\caption{Pauli spin susceptibility $\chi_P$ as a function of
temperature $T$ in the unit of hopping parameter $t$ for $N_s=
8^3$ and $k_F = 30$ MeV. The vertical dashed line signifies
$T^\ast (N_s=8^3) = 0.5253(3) t$, which is determined as the
maximum point of $\chi_P(T)$, using the fitting function
Eq.~(\ref{magsm}) with $C_1 = 0.258 \pm 0.008$ (shown as the solid
curve).} \label{Fig:chiP8_kF30}
\end{center}
\end{figure}

Figure~\ref{Fig:chiP8_kF30} shows a typical case.  For $k_F = 30$
MeV and $N = 8^3$, we obtain $T^\ast /t = 0.5253(3)$ 
[$T^\ast = 0.06624(3)$ MeV with $t = 0.1261$ MeV]. 
The maximum point of the
Monte Carlo data is determined through interpolation by use of a
fitting function with a parameter $C_1$,
\begin{equation}
\chi_P (T) = C_1 T \exp( - T/T^\ast).
\label{magsm}
\end{equation}

Note that though the definition of $T^\ast$ is somewhat
subjective, $T^\ast$ thus defined approaches $T_c$ at the BCS
limit, and $T^\ast$ signifies the pair-forming temperature at the
BEC limit as $T^\ast \propto |c_0| / \left[ a^3t
\ln(|c_0|/(a^3t\epsilon_F))^{3/2} \right]$ \cite{Sewer,Randeria}.  Here,
the BCS and BEC limits correspond to the weak and strong
interaction limits, or the small and large $c_0/(a^3t)$ limits,
respectively.

\section{$\Delta$, $T_c$, and $T^\ast$
at the Thermodynamic and continuum limits}

\subsection{Pairing gap $\Delta$}

\begin{figure}[htbp]
\begin{center}
\includegraphics[width=100mm]{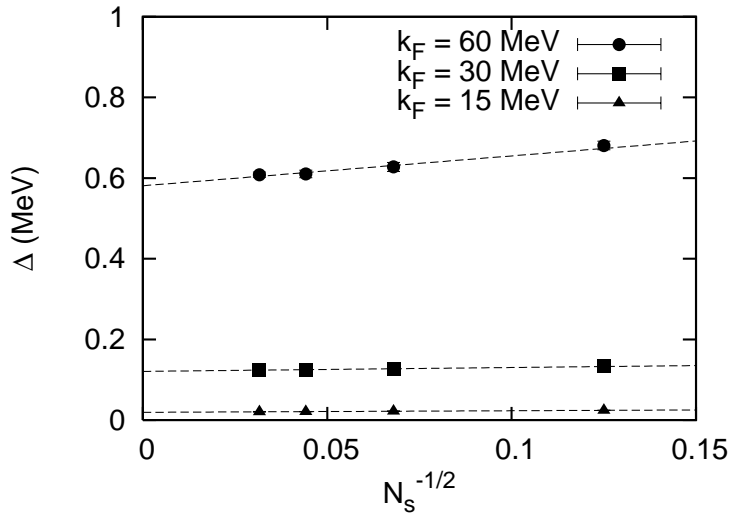}
\caption{Lattice-size ($N_s$) dependence of $\Delta$ at LO for
$k_F = 15$, $30$, and $60$ MeV with $n = 1/4$. The Monte Carlo
data shown with the statistical uncertainties are obtained
for $Ns = 4^3$, $6^3$, $8^3$, and $10^3$. The dashed lines are
the best fits by the use of linear functions of $N_s^{-1/2}$.}
\label{Fig:Delta_FSS_LO}
\end{center}
\end{figure}

\begin{figure}[htbp]
\begin{center}
\includegraphics[width=100mm]{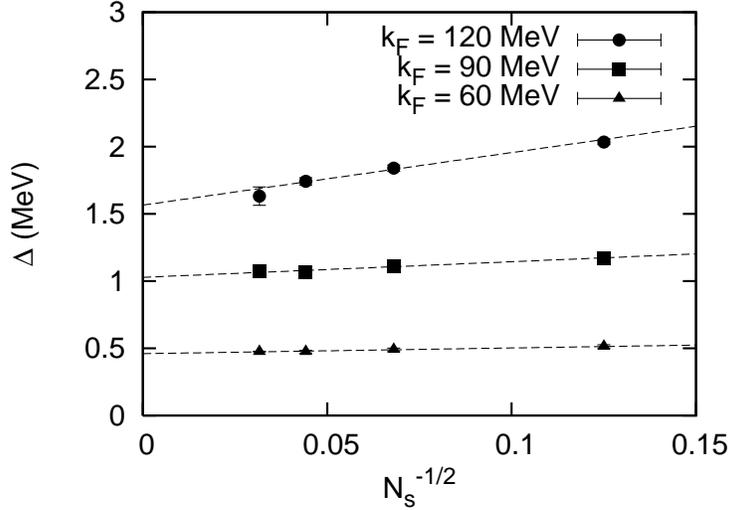}
\caption{Same as Fig.~\ref{Fig:Delta_FSS_LO}, but at NLO
for $k_F = 60$, $90$, and $120$ MeV.}
\label{Fig:Delta_FSS_NLO}
\end{center}
\end{figure}

As the first step, we determine $\Delta$ at the thermodynamic
limit.  To carry out a definite analysis, we apply
the BCS finite-size scaling exponent, $\lambda = 3/2$ in $\Delta
\sim L^{-\lambda} = N_s^{-\lambda/3}$ as being independent of the
density \cite{Schneider}. The exponent is obtained through
$\Delta(T=0, N_s) \propto T_c(N_s) \sim L^{-3/2} = N_s^{-1/2}$ by
combining the BCS result, $\Delta(T=0) \approx 1.76 T_c$, and the
direct relations between the finite-size scaling and critical
exponents \cite{Schneider,fscaling}.  Note that while the usual
$\chi^2$ best fit to all of our Monte Carlo data results in an
essentially indefinite $\lambda$, the jackknife method (often used
in the lattice QCD data analysis \cite{LatticeQCD}) yields
$\lambda = 1.6 \pm 0.3$ by assuming a linear $L^{-\lambda}$
dependence independent of the density. Apparently the value of the
exponent changes little between the BCS weak-coupling region and
the neutron-matter BCS-BEC crossover region.  Figures
~\ref{Fig:Delta_FSS_LO} and \ref{Fig:Delta_FSS_NLO} show the
choice of $\lambda = 3/2$ reasonable.

In Figs.~\ref{Fig:Delta_FSS_LO} and \ref{Fig:Delta_FSS_NLO},
the finite-size scaling of $\Delta$ is shown as a function of
$N_s$ using $N_s = 4^3$, $6^3$, $8^3$, and $10^3$ data with $n =
1/4$. Fig.~\ref{Fig:Delta_FSS_LO} is the finite-size scaling of
$\Delta$ evaluated at LO for $k_F = 15$, $30$, and $60$ MeV, while
Fig.~\ref{Fig:Delta_FSS_NLO} is at NLO for $k_F = 60$, $90$, and
$120$ MeV. The data at LO are found to be best fit with a linear
dependence on $N_s^{-1/2}= L^{-3/2}$ as
\begin{eqnarray}
\label{Eq:Delta_FSS_LO}
  \Delta (N_s, k_F = 15 \ {\rm MeV})
  &=& 0.0394(34) \ N_s^{-1/2} + 0.019152(20), \nonumber \\
  \Delta (N_s, k_F = 30 \ {\rm MeV})
  &=& 0.096(35) \ N_s^{-1/2} + 0.1207(16), \\
  \Delta (N_s, k_F = 60 \ {\rm MeV})
  &=& 0.74(24) \ N_s^{-1/2} + 0.581(13), \nonumber,
\end{eqnarray}
and those for NLO are
\begin{eqnarray}
\label{Eq:Delta_FSS_NLO}
  \Delta (N_s, k_F = 60 \ {\rm MeV})
  &=& 0.423(67) \ N_s^{-1/2} + 0.4602(54), \nonumber \\
  \Delta (N_s, k_F = 90 \ {\rm MeV})
  &=& 1.16(17) \ N_s^{-1/2} + 1.028(14), \\
  \Delta (N_s, k_F = 120 \ {\rm MeV})
  &=& 3.91(75) \ N_s^{-1/2} + 1.565(42), \nonumber
\end{eqnarray}
where the last constant for each value of $k_F$ is $\Delta$ at the
thermodynamic limit ($N_s \rightarrow \infty$).  The
best-fit constants in Eqs.~(\ref{Eq:Delta_FSS_LO}) and
(\ref{Eq:Delta_FSS_NLO}) are determined using the jackknife method.

\begin{figure}[htbp]
\begin{center}
\includegraphics[width=100mm]{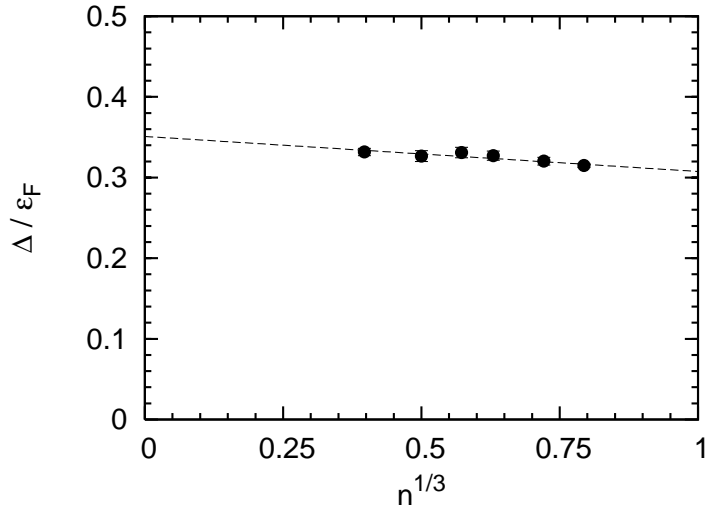}
\caption{Pairing gap $\Delta$ in the unit of the Fermi energy
$\epsilon_F$ as a function of the filling fraction $n$ for $k_F =
60$ MeV.  The solid circles show Monte Carlo data for $N_s =
6^3$ at LO, with statistical uncertainties. The dashed line is
the best fit by the use of a linear $n^{1/3}$ dependence. The
interception of the dashed line with the vertical axis corresponds
to $\Delta$ at the continuum limit ($n \rightarrow 0$) for $N_s =
6^3$ at $k_F = 60$ MeV.} \label{Fig:Delta_adep}
\end{center}
\end{figure}

As the second step, we determine $\Delta$ in the continuum
limit using the above thermodynamic limit values.  As discussed
in Sec.~II C, these values are obtained by using the
standard parameter set, or with $n = 1/4$ (half of the
quarter-filling), and are needed to extrapolate to $n = 0$ to
reach the continuum limit, $a \rightarrow 0$. For the
extrapolation, we need to know how much $\Delta$ changes between
$n =1/4$ and $n \rightarrow 0$, or the ratio of $\Delta$ at the
two values of $n$, $R_{\Delta}$. In this work, we determine
$R_{\Delta}$ solely using LO Monte Carlo data of the $N_s = 6^3$
lattice for $k_F = 60$ MeV. Dependence of $R_{\Delta}$ on $N_s$
and $k_F$ is weak both for LO and NLO, as discussed in
Sec.~IV D.

Figure~\ref{Fig:Delta_adep} shows the dependence of $\Delta$
on $n$ for $N_s = 6^3$ at $k_F = 60$ MeV. The data in the figure,
shown with statistical uncertainties by solid circles, are for $n
= 1/16$, $1/8$, $3/16$, $1/4$, $3/8$, and $1/2$.  The EFT
potential parameter $c_0(a)$ is varied by the use of
Eq.~(\ref{c00}) to accommodate the variation of $a$ generated by
the change of $n$. 

The $n$ dependence of $\Delta$ is found to be relatively weak, and the jackknife analysis of the data yields 
\begin{equation}
  \Delta(n, N_s = 6^3)/\epsilon_F = - 0.07(7) \ n^{1.6(1.3)} + 0.337(20).
\label{jknife-fit}
\end{equation}
While more data are desirable to reduce the uncertainty of the continuum limit, the constant term in Eq.~(\ref{jknife-fit}),  some indirect information of the $n$ exponent is available from the weak-coupling BCS theory by the use of $\Delta \propto T_c$, and also from the analysis  by Burovski {\it et al}. \cite{burovski} in a similar limit (but with $k_F \rightarrow 0$ as noted in Sec.~II C) for their unitary limit calculation.   Both suggest  the $n^{1/3}$ dependence, with which we find the best fit 
\begin{equation}
  \Delta(n, N_s = 6^3)/\epsilon_F = - 0.044(16) \ n^{1/3} + 0.351(10).
\label{deltacont}
\end{equation}
For definiteness and because of lack of time,  we use in our present analysis    Eq.~(\ref{deltacont}) and show it as the dashed line in Fig.~\ref{Fig:Delta_adep}.   Equations (\ref{jknife-fit}) and (\ref{deltacont}) yield the statistically consistent $\Delta$ at the continuum limit and suggest the systematic uncertainty by the use of the $n^{1/3}$ dependence to be several percent.

Equation~(\ref{deltacont}) gives
the ratio $R_{\Delta}$
\begin{equation}
  R_{\Delta}
  \equiv \frac{\Delta(n \rightarrow 0, N_s = 6^3)}{\Delta(n = 0.25, N_s = 6^3)}
  = \frac{0.674(19) \ {\rm MeV}}{0.628(11) \ {\rm MeV}}
  = 1.07(5).
\label{Eq:Delta_adep}
\end{equation}
That is, the continuum-limit correction amounts to a 7\%
increase in the value of $\Delta$.  Exploiting the weak
dependence of $R_{\Delta}$ on $N_s$ and $k_F$ (elaborated in
Sec.~IV D), we apply the same $R_{\Delta}$ to $\Delta$ at the
thermodynamic limit in Eqs.~(\ref{Eq:Delta_FSS_LO})
and (\ref{Eq:Delta_FSS_NLO}), so as
to obtain the final values of $\Delta$ at the thermodynamic and
continuum limits.

\subsection{Critical temperature $T_c$ and pairing temperature scale $T^\ast$}

\begin{figure}[htbp]
\begin{center}
\includegraphics[width=100mm]{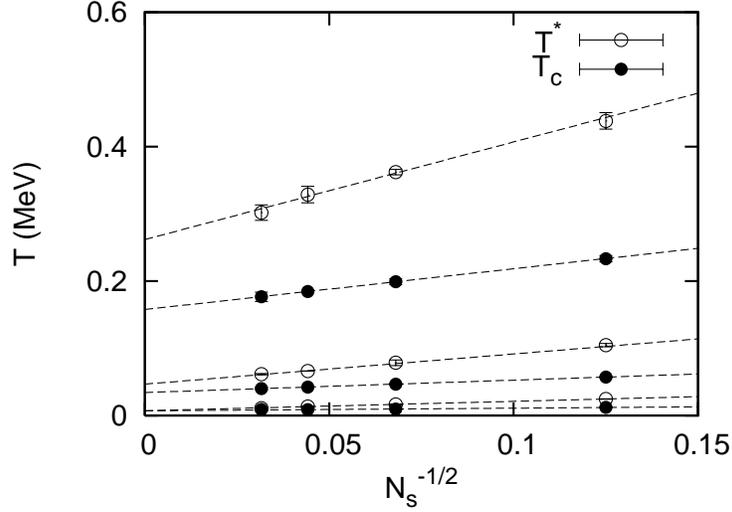}
\caption{Finite-size scaling of the critical temperature $T_c$
and the pairing temperature scale $T^\ast$. The Monte Carlo data
for $T_c$ and $T^\ast$ with $n = 1/4$ at LO are shown 
for $k_F = 15$, $30$, and $60$ MeV from
bottom to top. The dotted lines are the best fits of Eqs.~(\ref
{Eq:Tc_FSS_LO}) and (\ref{Eq:Tast_FSS_LO}).}
\label{Fig:T_FSS_LO}
\end{center}
\end{figure}

\begin{figure}[htbp]
\begin{center}
\includegraphics[width=100mm]{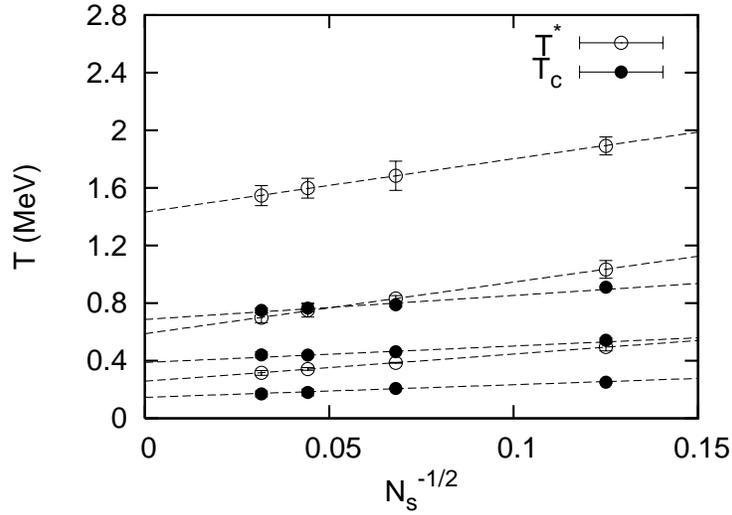}
\caption{Same as Fig.~\ref{Fig:T_FSS_LO}, 
for $k_F = 60$, $90$, and $120$ MeV, from bottom to top.
The dotted lines are the best fits of Eqs.~(\ref{Eq:Tc_FSS_NLO})
and (\ref{Eq:Tast_FSS_NLO}).}
\label{Fig:T_FSS_NLO}
\end{center}
\end{figure}

To obtain $T_c$ and $T^\ast$ at the thermodynamic and
continuum limits, we carry out the same two steps as those done on
$\Delta$ in the preceding subsection.  Because $T_c$ is at
criticality, we will apply the universality argument for taking
the thermodynamic limit.  Monte Carlo data at $n = 1/4$ used for
the finite-size scaling of $T_c$ and $T^\ast$ are shown in
Figs.~\ref{Fig:T_FSS_LO} and \ref{Fig:T_FSS_NLO} for $N_s = 4^3$,
$6^3$, $8^3$, and $10^3$ with statistical uncertainties.

The exponent of the finite-size scaling and the critical exponents
are known to be directly related at criticality \cite{fscaling}.
Furthermore, because the three-dimensional (3D) XY model and
our 3D Hubbard model are expected to belong to the same
universality class \cite{Sewer,Engelbrecht:2001,Schneider}, the
exponents of finite-size scaling at criticality of both models are
also expected to be the same \cite{Schneider, fscaling}.
Accordingly, we have $T_c(k_F, N_s) - T_c(k_F, N_s \rightarrow
\infty) \sim N_s^{-1/(3\nu)} = L^{-1/\nu}$ with $\nu = 2/3$ of the
XY model \cite{Engelbrecht:2001, critical_phenomena_reviews}.
Here, $\nu$ denotes the exponent of, for example, the correlation
length, as $\sim (T-T_c)^{-\nu}$. Note that in comparison, a
mean-field approximation such as Ginzburg-Landau theory gives $\nu
= 1/2$ \cite{exponent_MF}. With the linear $N_s^{-1/2}$
dependence, we find the best fits to the data for $T_c$ at LO to be
\begin{eqnarray}
\label{Eq:Tc_FSS_LO}
  T_c (k_F = 15 \ {\rm MeV}, N_s)
  &=& 0.039(14) \ N_s^{-1/2} + 0.00700(94), \nonumber \\
  T_c (k_F = 30 \ {\rm MeV}, N_s)
  &=& 0.1839(11) \ N_s^{-1/2} + 0.03420(11), \\
  T_c (k_F = 60 \ {\rm MeV}, N_s)
  &=& 0.6069(64) \ N_s^{-1/2} + 0.15770(30), \nonumber
\end{eqnarray}
which are shown in Fig.~\ref{Fig:T_FSS_LO},
and at NLO,
\begin{eqnarray}
\label{Eq:Tc_FSS_NLO}
  T_c (k_F = 60 \ {\rm MeV}, N_s)
  &=& 0.88(19) \ N_s^{-1/2} + 0.146(13), \nonumber \\
  T_c (k_F = 90 \ {\rm MeV}, N_s)
  &=& 1.14(33) \ N_s^{-1/2} + 0.388(22), \\
  T_c (k_F = 120 \ {\rm MeV}, N_s)
  &=& 1.67(63) \ N_s^{-1/2} + 0.687(37), \nonumber
\end{eqnarray}
which are shown in Fig.~\ref{Fig:T_FSS_NLO}.
The last constant in each best fit in Eqs.~(\ref{Eq:Tc_FSS_LO})
and (\ref{Eq:Tc_FSS_NLO}) is $T_c$ at
the thermodynamic limit, $T_c(k_F, N_s \rightarrow
\infty)$.

While $T^\ast$ is not at criticality, we find the finite-size
scaling for $T^\ast$ to be similar to that of $T_c$. For example,
the data of $T^\ast(N_s)$ yield the best-fit scaling power $\sim
N_s^{-0.507 \pm 0.007}$ with the jackknife method (for $T_c$,
$\sim N_s^{-0.53 \pm 0.03}$). We thus apply the same linear
$N_s^{-1/2}$ dependence to $T^\ast$ as that for $T_c$. The best
fits for $T^\ast$ at LO are found to be
\begin{eqnarray}
\label{Eq:Tast_FSS_LO}
  T^\ast (k_F = 15 \ {\rm MeV}, N_s)
  &=& 0.1400(24) \ N_s^{-1/2} + 0.00707(24), \nonumber \\
  T^\ast (k_F = 30 \ {\rm MeV}, N_s)
  &=& 0.448(54) \ N_s^{-1/2} + 0.0463(23), \\
  T^\ast (k_F = 60 \ {\rm MeV}, N_s)
  &=& 1.45(15) \ N_s^{-1/2} + 0.2618(99), \nonumber
\end{eqnarray}
and at NLO,
\begin{eqnarray}
\label{Eq:Tast_FSS_NLO}
  T^\ast (k_F = 60 \ {\rm MeV}, N_s)
  &=& 1.890(59) \ N_s^{-1/2} + 0.2575(35), \nonumber \\
  T^\ast (k_F = 90 \ {\rm MeV}, N_s)
  &=& 3.583(60) \ N_s^{-1/2} + 0.5876(61), \\
  T^\ast (k_F = 120 \ {\rm MeV}, N_s)
  &=& 3.70(12) \ N_s^{-1/2} + 1.4320(69), \nonumber
\end{eqnarray}
where the last constant in each equation gives $T^\ast$ at the
thermodynamic limit.

\begin{figure}[htbp]
\begin{center}
\includegraphics[width=100mm]{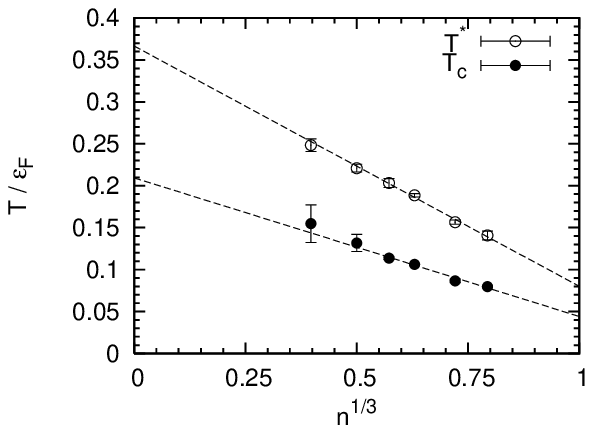}
\caption{$n$ dependence of the critical temperature $T_c$ and
pairing temperature scale $T^\ast$ at LO in the unit of the Fermi energy
$\epsilon_F$ for $N_s = 6^3$ at $k_F = 60$ MeV. The lines are 
the best fits to the $T_c$ and $T^\ast$ data, 
Eqs.~(\ref{Eq:T_c_adep}) and (\ref{Eq:Tast_adep}), respectively.}
\label{Fig:T_adep}
\end{center}
\end{figure}

As to the continuum limit, in Fig.~\ref{Fig:T_adep} we show the
$n$ dependence of $T_c$ and $T^\ast$ at LO for $N_s = 6^3$ at $k_F
= 60$ MeV. The data with statistical uncertainties are shown by
solid circles for $n = 1/16$, $1/8$, $3/16$, $1/4$, $3/8$, and
$1/2$. The exponent fit of $T_c$ ($T^\ast$) shows $T_c \sim
n^{0.31 \pm 0.12}$ ($T^\ast \sim n^{0.43 \pm 0.10}$). As observed
for the similar limit of $T_c$ \cite{burovski},
they appear to be best fit by a linear $n^{1/3}$ dependence,
\begin{equation}
  T_c(n, N_s = 6^3)/\epsilon_F = - 0.165(23) \ n^{1/3} + 0.209(12),
\label{Eq:T_c_adep}
\end{equation}
and
\begin{equation}
  T^\ast(n, N_s = 6^3)/\epsilon_F = - 0.286(20) \ n^{1/3} + 0.367(12).
\label{Eq:Tast_adep}
\end{equation}
Note that the continuum limits of $T_c$ and $T^\ast$ in
Eqs.~(\ref{Eq:T_c_adep}) and (\ref{Eq:Tast_adep}) are consistent
with those determined by the exponent fits using the
jackknife method within the statistical uncertainties [$T_c =
0.223(41)$ and $T^\ast = 0.328(34)$]. Contrary to the case of
$\Delta$, the $n$ dependence of $T_c$ and $T^\ast$ is rather
strong. Equations~(\ref{Eq:T_c_adep}) and (\ref{Eq:Tast_adep})
provide the needed ratios $R_{T_c}$ and $R_{T^\ast}$, which are
used to obtain $T_c$ and $T^\ast$ at the thermodynamic and
continuum limits, as in the case of $\Delta$.

\subsection{Dependence of the continuum limit on $N_s$ and $k_F$}

The extrapolation to $n \rightarrow 0$ depends generally on
$N_s$ and $k_F$, but the dependence is expected to be weak because
of the separation of local (ultraviolet) and global (infrared)
properties for a sufficiently large $N_s$.

For the $N_s$ dependence, we calculate, using the lattice sizes of
$N_s = 4^3$ and $8^3$, the ratios between $n \rightarrow 0$
and $n =0.25$: $R_{\Delta}$, $R_{T_c}$, and $R_{T^\ast}$, both at
LO and NLO. As summarized in Table \ref{Table:Dep_CL_Ns}, each
ratio at $k_F = 60$ MeV is consistent within the statistical
uncertainties for $N_s = 4^3$, $6^3$, and $8^3$ both at LO
and NLO. Note that the second row for $N_s = 6^3$ is obtained
using data at $n = 1/16$, $1/8$, $3/16$, $1/4$, $3/16$, and $1/2$,
while the other rows use data at $n = 1/16$, $1/4$, and
$1/2$.

Table \ref{Table:Dep_CL_kF} also confirms the weak dependence on
$k_F$. Note that the third row uses data for $n = 1/16$, $1/8$,
$3/16$, $1/4$, $3/8$, and $1/2$, while the other rows use
data at $n = 1/16$, $1/4$, and $1/2$.

\begin{table}[htbp]
\caption[Dependence of the continuum limit on $N_s$.]
{Dependence of the continuum limit on $N_s$.}
\label{Table:Dep_CL_Ns}
    \begin{center}
    \begin{tabular}{cccccc}
      \hline
      \hline
          & $k_F$ (MeV) & $N_s$ & $R_{\Delta}$  & $R_{T_c}$ & $R_{T^\ast}$ \\
      \hline
       LO & 60 & $4^3$ & \, 1.14(17) & \, 2.10(15) & 1.9(2) \\
       LO & 60 & $6^3$ & 1.07(5)  & \, 1.96(13) & \, 1.94(9) \\
       LO & 60 & $8^3$ & 1.12(8)  & \, 1.86(10) & 2.0(1) \\
      NLO & 60 & $4^3$ & 1.08(6)  & 2.08(8)  & 2.1(2) \\
      NLO & 60 & $6^3$ & 1.04(5)  & 2.05(7)  & 2.1(1) \\
      NLO & 60 & $8^3$ & 1.05(4)  & \, 2.08(37) & 2.0(1) \\
      \hline
      \hline
    \end{tabular}
  \end{center}
\end{table}

\begin{table}[htbp]
\caption[Dependence of the continuum limit on $k_F$.]
{Dependence of the continuum limit on $k_F$.}
\label{Table:Dep_CL_kF}
    \begin{center}
    \begin{tabular}{cccccc}
      \hline
      \hline
          & $k_F$ (MeV) & $N_s$ & $R_{\Delta}$  & $R_{T_c}$ & $R_{T^\ast}$ \\
      \hline
       LO & \, 15 & $6^3$ & 1.09(3)  & 1.96(9)  & 2.4(9)  \\
       LO & \, 30 & $6^3$ & 1.08(8)  & 1.98(6)  & 2.4(4)  \\
       LO & \, 60 & $6^3$ & 1.07(5)  & \, 1.96(13) & \, 1.94(9) \\
      NLO & \, 60 & $6^3$ & 1.04(5)  & 2.05(7)  & 2.1(1)  \\
      NLO & \, 90 & $6^3$ & \, 1.11(10) & \, 2.12(20) & 2.0(1)  \\
      NLO & 120 & $6^3$ & 1.04(2)  & \, 2.00(36) & 2.0(3)  \\
      \hline
      \hline
    \end{tabular}
  \end{center}
\end{table}

\section{Matching LO and NLO results}

\begin{figure}[htbp]
\begin{center}
\includegraphics[width=100mm]{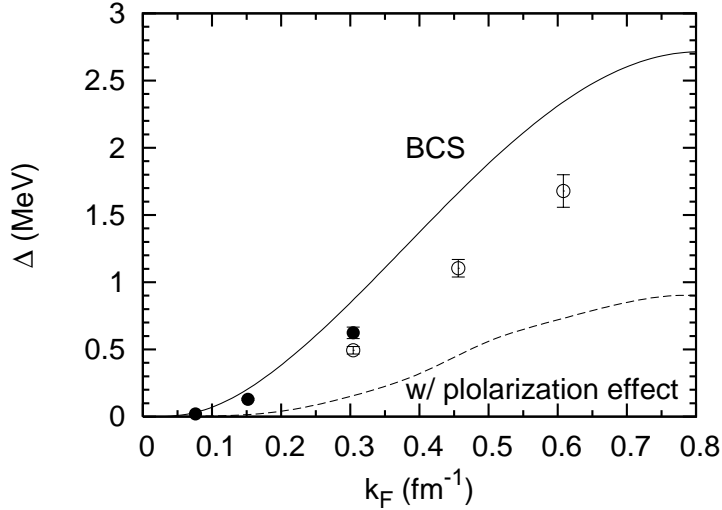}
\caption{$^1S_0$ pairing gap, $\Delta$, in the thermodynamic
and continuum limits, resulting from the LO 
(solid circles) and NLO (open circles) calculations. 
The neutron density is denoted in terms of the
Fermi momentum $k_F$. The BCS calculation of
Ref.~\cite{Elgaroy:1996mx} (solid curve) and a higher order
calculation including polarization effects of
Ref.~\cite{Wambach:1992ik} (dashed curve) are also shown for
comparison.  For a more detailed comparison, see
Fig.~\ref{Fig:Delta_comparison} in
Sec.~\ref{Sec:Delta_comparison}.} \label{Fig:Delta_kF.fss}
\end{center}
\end{figure}

\begin{figure}[htbp]
\begin{center}
\includegraphics[width=100mm]{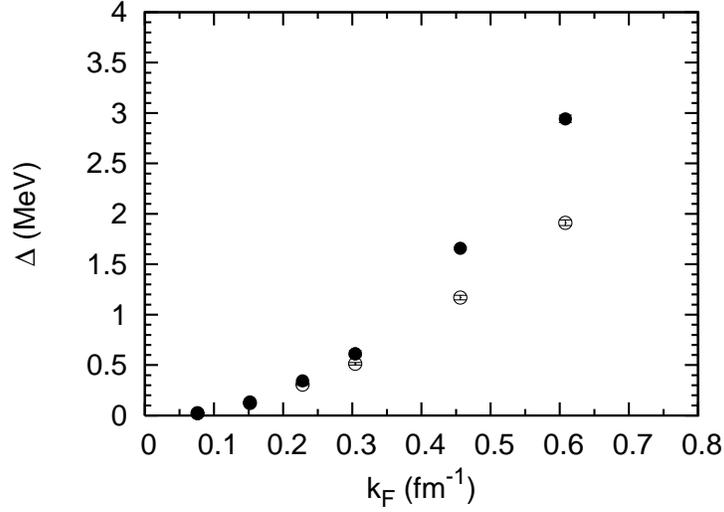}
\caption{$^1S_0$ pairing gap, $\Delta$, for $N_s = 4^3$ and $n
= 1/4$, resulting from the LO (solid circles) 
and NLO (open circles) calculations.
The neutron density is denoted in terms of the Fermi momentum
$k_F$.} \label{Fig:Delta_kF_matching}
\end{center}
\end{figure}

Figures~\ref{Fig:Delta_kF.fss} and \ref{Fig:Delta_kF_matching}
display the LO and NLO $\Delta$'s as a function of $k_F$ and
illustrate their matching in the region of $k_F = 0.15$-$0.30
\;{\rm fm}^{-1}$:  the $\Delta$ shown in
Fig.~\ref{Fig:Delta_kF.fss} is the result of the elaborate
calculation described in Secs.~III and IV, while the
$\Delta$ shown in Fig.~\ref{Fig:Delta_kF_matching} is the result
of a simpler calculation for $4^3$ lattices with $n = 1/4$,
including $\Delta$ at the density of $k_F = 0.22805$.  The density
dependences of the $\Delta$'s are quite close to each other in the two
figures, demonstrating that a smooth transition from the LO
$\Delta$ to the NLO $\Delta$ occurs in the density region of $k_F
= 0.15$-$0.30 \;{\rm fm}^{-1}$.  Accordingly, we take the LO
$\Delta$ for $k_F = 0.1520\;{\rm fm}^{-1}$ and the NLO $\Delta$
for $k_F = 0.3041\;{\rm fm}^{-1}$, as the final values.

\begin{figure}[htbp]
\begin{center}
\includegraphics[width=100mm]{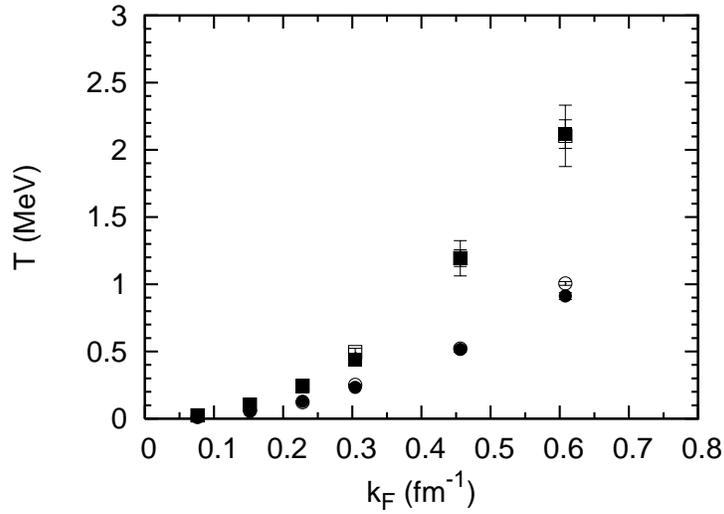}
\caption{Critical temperature $T_c$ (circles) and the pairing
temperature scale $T^\ast$ (squares) by the LO (solid symbols) 
and NLO (open symbols) calculations for $N_s
= 4^3$ and $n = 1/4$, shown as a function of the neutron matter
density (represented by the Fermi momentum $k_F$) 
The error bars are statistical uncertainties
only.} \label{Fig:phase_diagram_matching}
\end{center}
\end{figure}

Figure~\ref{Fig:phase_diagram_matching} shows that also for
$T_c$ and $T^\ast$, smooth transitions take place between the LO
and NLO values in the same density region as for $\Delta$.  We
thus also take $T_c$ and $T^\ast$ at $k_F = 0.1520\;{\rm fm}^{-1}$
as the LO and $T_c$ and $T^\ast$ at $k_F = 0.3041\;{\rm fm}^{-1}$
as the NLO. Note that the difference between the LO and NLO values
of $T_c$ and $T^\ast$ in Fig.~\ref{Fig:phase_diagram_matching} is
much smaller than that in the case of $\Delta$.

\section{Results}

\subsection{Pairing gap $\Delta$} \label{delta}

\begin{table}[h]
\caption{Our final values of the $^1S_0$ pairing gap $\Delta$ in
the thermodynamic and continuum limits, and the ratio of
$\Delta$ and the BCS value $\Delta_{\rm BCS}$.  Uncertainties 
are statistical only.} \label{table:Delta}
    \begin{center}
    \begin{tabular}{cccc}
      \hline
      \hline
      $k_F$ (MeV) & $\rho$ ($\rho_0$) & $\Delta$ (MeV)& $\Delta/\Delta_{\rm BCS}$\\
      \hline
       \, 15 & $9 \times 10^{-5}$ & \, \, 0.021(1) & 0.69(3)\\
       \, 30 & $7 \times 10^{-4}$ & \, 0.13(1)  & 0.67(4)\\
       \, 60 & $6 \times 10^{-3}$ & \, 0.49(3)  & 0.56(5)\\
       \, 90 & $2 \times 10^{-2}$ & \, 1.10(7)  & 0.68(4)\\
      120 & $5 \times 10^{-2}$ & 1.7(1)   & 0.74(4)\\
      \hline
      \hline
    \end{tabular}
  \end{center}
\end{table}

Table \ref{table:Delta} lists our final values of $\Delta$ in the
thermodynamic and continuum limits for low-density neutron
matter. Table \ref{table:Delta} includes the ratio of $\Delta$ and
the corresponding BCS pairing gap, $\Delta_{\rm BCS}$. Here, the
$\Delta_{\rm BCS}$'s are taken from those tabulated in
Ref.~\cite{Elgaroy:1996mx} as the representative BCS values.  As noted
in Sec.~VII B, there are only quite small differences among
the $\Delta_{\rm BCS}$'s calculated by the CD-Bonn, Nijmegen I,
Nijmegen II, and Argonne V18 $NN$ potentials
\cite{Elgaroy-1998,schwenk-2007}.

It is difficult to assess the systematic uncertainties
involved in our calculation.  In view of the probable
uncertainties involved in taking the thermodynamic limit and
especially the continuum limit, however, it would be fair to
state that our calculation yields $\Delta$ to be approximately
30\% less than the BCS values, perhaps with an additional
systematic uncertainty of about $\pm$10\%.  We thus
consider finer variations of $\Delta$ inconclusive.  For example,
a close examination of Table \ref{table:Delta} shows that the
$\Delta/\Delta_{\rm BCS}$ ratio dips at around $k_F = 60$ MeV.
But this would require further study.

\subsection{Phase diagram of low-Density neutron matter}

Table~\ref{table:DeltaTc} lists our final values of $T_c$ and
$T^\ast$ in the thermodynamic and continuum limits.  It also
shows their ratios and the ratios with the $\Delta$ of
Table~\ref{table:Delta}.  In Table~\ref{table:DeltaTc}, we observe
that $T^\ast$ approaches $T_c$ as the density decreases.
That is, the pseudogap state (see below) diminishes as the
density decreases.  Furthermore, as the density decreases, the
$\Delta /T_c$ ratio approaches the BCS value of about 1.76
\cite{Schrieffer}, while $\Delta$ and $T_c$ themselves remain
different from the BCS values.

\begin{figure}[htbp]
\begin{center}
\includegraphics[width=100mm]{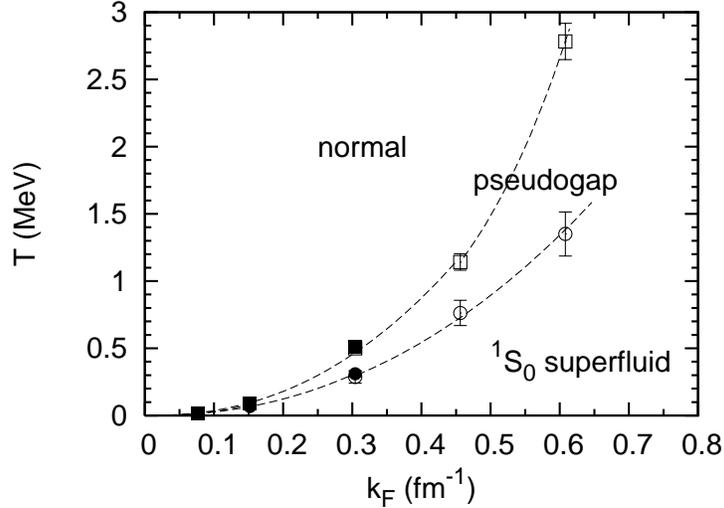}
\caption{${}^1S_0$ phase diagram of low-density neutron
matter. The solid and open symbols with statistical uncertainties
show the LO and NLO results, respectively. The dotted curves for
$T_c$ and $T^\ast$ are drawn by extrapolation. Neutron matter is
in the superfluid phase below the critical temperature $T_c$ of
the second-order phase transition. Above $T_c$, neutron matter is
in the pseudogap phase \cite{Schneider}, in which pairing remains
locally without forming long-range order, and undergoes a smooth
transition from the pseudogap phase to the normal phase around
$T^\ast$, as pairing gets much less.}
\label{Fig:phase_diagram_fss}
\end{center}
\end{figure}

$T_c$ and $T^\ast$ in Table~\ref{table:DeltaTc} provide the
temperature-density phase diagram as shown in
Fig.~\ref{Fig:phase_diagram_fss}.  The figure illustrates the
thermodynamic properties of low-density neutron matter.
For example, at a fixed density $k_F$, as the temperature goes
down from the normal phase, the pairing is gradually enhanced,
forming the pseudogap phase \cite{Schneider}  around and below
$T^\ast$. As the temperature goes down farther, the pairing gets
stronger and eventually forms a long-range ordering at $T_c$,
thereby generating the second-order phase transition to the
superfluid phase. Note that the transition between the pseudogap
phase and the normal phase is smooth.  We must also note that the
definition of $T^\ast$ is somewhat subjective.

\begin{table}[htbp]
\caption{Our final values of $T_c$ and $T^\ast$, and the relative
magnitudes among them and $\Delta$ in Table \ref{table:Delta}.}
\label{table:DeltaTc}
    \begin{center}
    \begin{tabular}{cccccc}
      \hline
      \hline
      $k_F$ (MeV) & $T_c$ (MeV) & $T^\ast$ (MeV) & $\Delta/T_c$
      & $\Delta/T^\ast$ & $T_c / T^\ast$ \\
      \hline
      \, 15 & \, \, 0.014(3)  & \, \, 0.014(1) & 1.5(4) & 1.5(2)   & \, 0.99(28)\\
      \, 30 & \, \, 0.067(5)  & \, \, 0.091(9) & 1.6(2) & 1.4(2)   & \, 0.74(12)\\
      \, 60 & \, 0.29(5)   & \, 0.45(5)  & 1.7(4) & \, \, 0.99(11) & \, 0.57(12)\\
      \, 90 & \, 0.76(9)   & 1.1(1)   & 1.5(3) & \, \, 0.97(11) & \, 0.67(12)\\
     120 & 1.4(2)    & 2.8(1)   & 1.2(2) & \, 0.60(7)  & 0.49(8)\\
      \hline
      \hline
    \end{tabular}
  \end{center}
\end{table}

\section{Discussions}
\label{discussions}

\subsection{Nature of low-density neutron matter: BCS-BEC crossover}

\begin{figure}[htbp]
\begin{center}
\includegraphics[width=100mm]{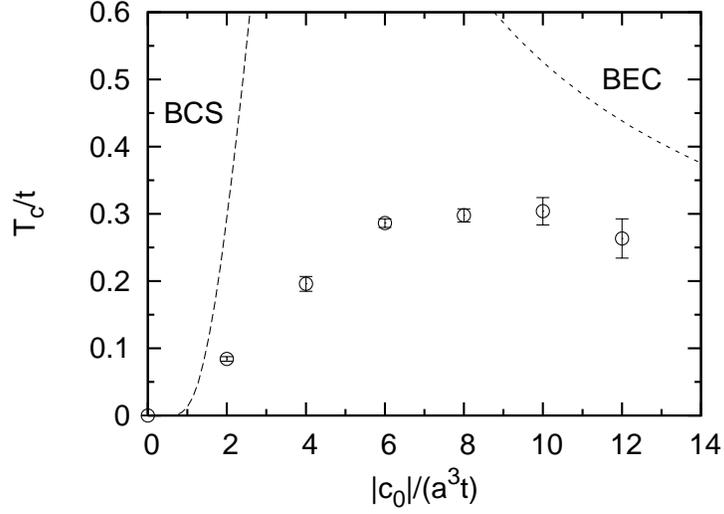}
\caption{EFT parameter ($c_0$) dependence of the critical
temperature $T_c$. For easier comparison, $T_c$ and $c_0$ are
expressed as dimensionless by use of the spatial lattice
spacing $a$ and the hopping parameter $t$. The open circles
are shown for $N_s = 6^3$ at the quarter-filling ($n = 0.5$).
The dashed curves are $T_c/t$ at the BCS and BEC limits of
Eq.~(\ref{tcurve}).} \label{Fig:BCS_BEC_crossover}
\end{center}
\end{figure}

To understand the nature of low-density neutron matter,
we examine the dependence of $T_c$ on the parameter $c_0$ by
applying the LO calculation, since the physics throughout
our low-density region is largely dictated by $c_0$.
Figure~\ref{Fig:BCS_BEC_crossover} illustrates the dependence in
comparison to $T_c$ in the weak-coupling (BCS) and strong-coupling
(BEC) limits,
\begin{eqnarray}
\label{tcurve}
 T_c({\rm BCS}) &=& \frac{2e^\gamma}{\pi}
 \sqrt{\left( 36t^2 - \mu^2 \right)}
 \exp\left( -\frac{a^3}{D_0(\mu) |c_0|}\right), \nonumber\\
 \\
 T_c({\rm BEC}) &=& 2 \left( \frac{2\pi^2 n}{\Gamma (3/2) \zeta (3/2)} \right)^{2/3}
 \frac{a^3t^2}{|c_0|}, \nonumber
\end{eqnarray}
respectively \cite{Sewer:PhDthesis}.  Here, $\gamma$ is Euler's
constant and $D_0(\mu)$ is the density of states.   In our
low-density neutron matter, $|c_0|/(a^3 t)$ is $5$-$7$, and
corresponds to the middle region in
Fig.~\ref{Fig:BCS_BEC_crossover}. The figure clearly shows that
{\it the thermal property of low-density neutron matter is not in
a state of BCS, but of BCS-BEC crossover.}  Though not discussed
here, the $c_0$ dependence of $T^\ast$ also verifies this point
\cite{Sewer, Sewer:PhDthesis}.

\begin{figure}[htbp]
\begin{center}
\includegraphics[width=100mm]{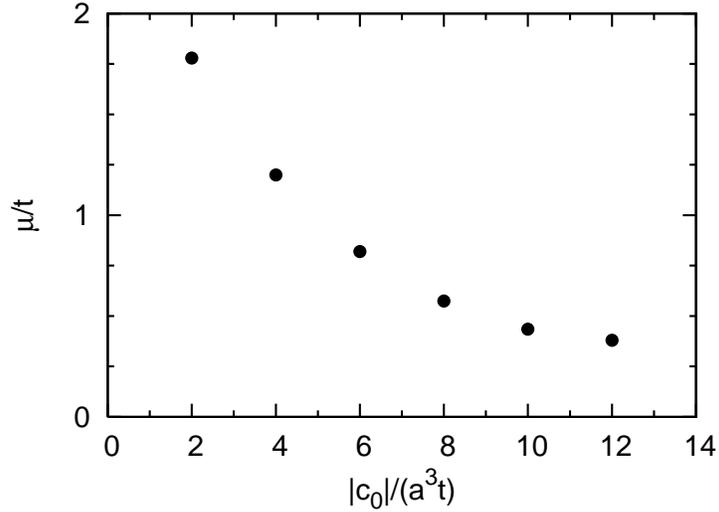}
\caption{Chemical potential $\mu$ as a function of the
interaction strength $c_0$ in a dimensionless unit, with the
spatial lattice spacing $a$ and the hopping amplitude $t$.  The
calculation is of the LO for $N_s=6^3$ and $n=0.5$.}
\label{Fig:chemical_potential}
\end{center}
\end{figure}

The preceding point is perhaps better clarified by the $c_0$
dependence of the chemical potential $\mu$.  $\mu$ is positive in
the weak-coupling BCS region and becomes negative in the
strong-coupling BEC region by exhibiting a bosonic nature.
Figure~\ref{Fig:chemical_potential} illustrates the $c_0$
dependence of $\mu$ in the LO calculation. $\mu$ decreases as
$c_0$ increases, and it takes a relatively small, positive value in
the region of our low-density neutron matter.  The small positive
value is in accord with the neutron matter being close but not
(yet) in the BEC region and indeed confirms the simple
characterization of the crossover, a negative and small (in
magnitude) value of $1/(k_F a_0)$ \cite{Randeria}, as noted in
Sec.~I.

\subsection{Pairing gap $\Delta$}
\label{Sec:Delta_comparison}

\begin{figure}[htbp]
\begin{center}
\includegraphics[width=100mm]{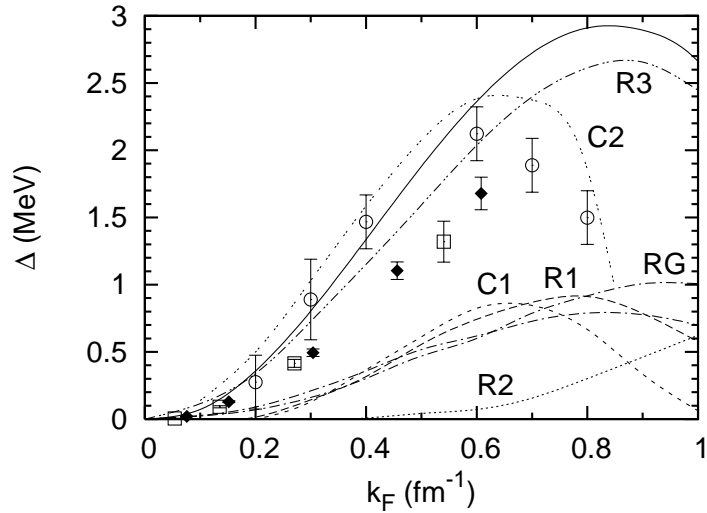}
\caption{Comparison of our Monte Carlo $\Delta$ to other
calculations as a function of the neutron matter density
(represented by the Fermi momentum $k_F$).  The solid
diamonds show our results, with statistical uncertainties. The
other calculations consist of three types: quantum Monte Carlo
(symbols with statistical uncertainties), BCS (solid curve), 
and BCS with higher-order effects (R's, C's, and
RG; shown by dotted and dashed curves). See text for the
description of each calculation shown.}
\label{Fig:Delta_comparison}
\end{center}
\end{figure}

Figure \ref{Fig:Delta_comparison} illustrates the density
dependence of various $\Delta$'s reported in the literature.
$\Delta$'s in the figure consist of those obtained by three types
of calculations: (1) BCS (shown by a solid curve), (2) BCS or
similar approximations, with higher order effects (dotted 
and dashed curves), and (3) quantum Monte Carlo (shown with
error bar symbols).

(1) Below $k_F \approx 0.7 \;{\rm fm}^{-1}$, there are few
recognizable differences \cite{Elgaroy-1998,schwenk-2007} among
$\Delta_{\rm BCS}$'s calculated by various conventional $NN$
potentials: Argonne $v_{18}$ \cite{Wiringa:1994wb}, Nijmegen
\cite{Stoks:1994wp}, and CD Bonn \cite{Machleidt:1995km}.
Accordingly, $\Delta_{\rm BCS}$'s are represented by a single (solid)
curve in Fig.~\ref{Fig:Delta_comparison}.

(2) Figure~\ref{Fig:Delta_comparison} includes $\Delta$'s by the
calculations beyond BCS.  Calculations in the random phase
approximation (RPA) with polarization effects are by Wambach
{\it et al.} \cite{Wambach:1992ik} (denoted as R1), by Schulze {\it et
al.}~\cite{Schulze:1996} (R2), and by Cao {\it et al.}~\cite{Cao:2006}
(R3). Calculations using correlated-basis functions are by Chen {\it et
al.}~\cite{Chen:1993} (C1) and by Fabrocini {\it et
al.}~\cite{Fabrocini:2005} (C2). A calculation based on a
renormalization group approach is by Schwenk {\it et
al.}~\cite{Schwenk:2003} (RG). The curves for these $\Delta$'s are
taken from similar figures in the recent literature: Figs.~1
and 2 of Ref.~\cite{Gandolfi:2008} and Fig.~4 of
Ref.~\cite{Gezerlis:2007}. In addition, though not shown, an
extrapolation from finite nuclei results obtained by
Hartree-Fock-Bogoliubov calculations also gives $\Delta$
close to the $\Delta_{\rm BCS}$ for $k_F \alt 0.5 \; {\rm fm}^{-1}$
~\cite{Margueron:2007}. We see that these $\Delta$'s differ
appreciably among each other, though recent works tend to give the
values closer to the BCS $\Delta$.

(3) Two types of quantum Monte Carlo calculations have been
reported based on the GFMC \cite{Carlson:2007, Gezerlis:2007} and
AFDMC \cite{Fabrocini:2005, Gandolfi:2008} methods. The two
methods are applied for a fixed number of neutrons using the
conventional $NN$ potentials (or some model potentials), while our
work is based on a grand canonical ensemble formulation. Figure
\ref{Fig:Delta_comparison} shows the most recent results of the
GFMC \cite{Gezerlis:2007} (open squares), the AFDMC
\cite{Gandolfi:2008} (open circles), and ours (taken from
Table \ref{table:Delta} and shown by solid diamonds).

In the figure, we see that all quantum Monte Carlo calculations
are, overall, close to the $\Delta_{\rm BCS}$.   The AFDMC $\Delta$
is quite close to the $\Delta_{\rm BCS}$ in the density region
examined in this work, while the GFMC $\Delta$ is smaller than the
$\Delta_{\rm BCS}$ and is similar to (even slightly lower than) our
$\Delta$.  Note that above $k_F \approx 0.6$ fm$^{-1}$, the AFDMC
$\Delta$ becomes quickly smaller than the $\Delta_{\rm BCS}$ as the
density increases.

It is difficult to assess the three quantum Monte Carlo
calculations by comparing them because the intermediate steps of
the calculations are all different.  Here, however, we point out a
possible issue closely tied to their basic formulations and
setups: stemming from the neutron numbers being fixed, the GFMC
and AFDMC $\Delta$'s are calculated using the odd-even staggering
(or the second-order finite difference) of the energy per neutron,
\begin{equation}
  \Delta ({\rm odd} \ N) = E(N) - \frac{1}{2}
  \left[ E(N-1) + E(N+1) \right],
\label{Eq:Deltaoddeven}
\end{equation}
where $N$ is the number of neutrons.  As described in Sec.~III,
our $\Delta$'s are calculated directly from the spin pair-pair
correlation functions.  By physical arguments, the two ways
of calculating $\Delta$ are expected to be the same for a large
$N$, but we are not aware of a rigorous proof for this
expectation. Since it has been a common practice to apply
Eq.~(\ref{Eq:Deltaoddeven}) for the extraction of $\Delta$ from
finite nuclei \cite{Dean:2002zx,abook}, closer examination of
this issue would be desirable, as exemplified in
Ref.~\cite{Margueron:2007}.

As noted above, it is desirable to apply
Eq.~(\ref{Eq:Deltaoddeven}) for a large $N$.  The large values up
to $N = 92$ are used in the GFMC calculation \cite{Gezerlis:2007},
while up to $N=68$ in the AFDMC \cite{Gandolfi:2008}.  Both $N$'s
are perhaps large enough to provide reliable information for
$N \rightarrow \infty$. While it might be caused by the
different ways the nuclear potentials are applied in the two
methods, the noticeable difference between the GFMC and AFDMC
$\Delta$'s is puzzling to us.

\subsection{Further improvement of the present work}

We note here the aspects of this work that we would like to
improve.

(1) The largest lattice size we have used is $N_s = 10^3$,
but larger lattices would be desirable for reliably reaching the
thermodynamic limit.  For this, we would like to study more
closely the use of the hybrid Monte Carlo (HMC) method. As
the commonly used method in lattice QCD calculations
\cite{LatticeQCD}, the HMC is expected to reduce the
computation time from $\sim (N_s N_t)^2$ or $(N_s N_t)^3$ (for the
DQMC) to $\sim (N_s N_t)^{5/4}$. Our trial application of the HMC
(following Ref.~\cite{Scalettar:1987}) in our problem has shown a
strong dependence on the HMC parameters, such as the size and
number of molecular dynamics steps and has brought about a
difficult compromise between the computation time and the
systematic error. We suspect that the difficulty stems from
badly conditioned fermion matrices and also from our
(effectively) strong interaction.  We would like to resolve this
issue and find a practical procedure for optimizing the HMC
calculation for this problem.

(2) Because of lack of time, we have examined the continuum limit
by applying the case of $N_s=6^3$ to all $N_s$'s that we computed.
The possible $N_s$ dependence is a potentially important source of
the systematic error, and we would like to clarify this issue.

(3) The matching of the LO and NLO calculations indicates that
our $\Delta$ deviates from the $\Delta_{\rm BCS}$ more appreciably in
the matching density region, $k_F \approx 0.15$-$0.3 \;{\rm
fm}^{-1}$. It is difficult to establish the deviation by
using the present statistics. We would like to examine this
density region more closely to determine whether such a fine
structure of the density dependence of $\Delta$ exists.

\section{Summary}

In conclusion, we have investigated thermal properties of
low-density neutron matter by the determinantal quantum Monte
Carlo lattice calculations with the single- and two-parameter
pionless EFT $NN$ potential.  The $^1S_0$ pairing gap at $T \approx
0$, the critical temperature of normal-to-superfluid phase
transition, and the pairing temperature scale have been
determined directly from the correlation functions and have
provided the temperature-density phase diagram for the density of
$(10^{-4}$-$10^{-1})\rho_0$. The thermodynamic limit was
taken, and the continuum limit was examined in the determination. The
pairing gap was found to be approximately 30\% less than the
BCS value. The physics of neutron matter in this density region
has clearly been identified as a BCS-BEC crossover.

\bigskip
\bigskip
\centerline{\bf ACKNOWLEDGMENTS}
\bigskip

We thank U. van Kolck for his continuing support for our project
by clarifying various aspects and issues on EFT, especially those
associated with the power counting rules and regularization
procedure. We acknowledge H.~M.~M{\"u}ller for allowing our use
and modification of his code, D. Lee for his useful comments after
reading the initial version of the manuscript, and K.-F. Liu and
T. Onogi for their instructive comments on our lattice
calculations. 
The calculations were carried out on Seaborg, Bassi, and Franklin at
the National Energy Research Scientific Computing Center, which is
supported by the Office of Science of the U.S.~Department of
Energy under Contract No.~DE-AC03-76SF00098, and at Titech Grid
and TSUBAME, Tokyo Institute of Technology, Japan. The major part
of this work was carried out at Kellogg Radiation Laboratory,
Caltech. We thank R.~McKeown for his generous hospitality over the
years. A part of this work was also performed at the Yukawa Institute for Theoretical Physics (YITP), Kyoto
University. R.~S.~is grateful for the warm hospitality received at the YITP. 
This work is supported by the U.S.~Department of
Energy under Grant No.~DE-FG02-87ER40347 at CSUN.
 \vfill\eject

\appendix

\section{Determination of the EFT potential parameters $c_0(\Lambda)$ and $c_2(\Lambda)$}
\label{eft-c0c2}

The EFT potential parameters, $c_0(\Lambda)$ and $c_2(\Lambda)$,
are determined from the observables for an appropriately chosen
value of $\Lambda$.  As the observables, we choose the
scattering length $a_0$ and the effective range $r_0$ in the
effective range expansion of Eq.~(\ref{effexp}) with $\Lambda =
\pi/a$ in our lattice calculation (where $a$ is the lattice spacing).

$\Lambda$ is needed in the determination of $c_0(\Lambda)$
and $c_2(\Lambda)$ so as to regularize loop contributions, which
otherwise diverge. With the regularization, the Schr\"{o}dinger
equation is solved, and $a_0$ and $r_0$ are expressed in terms of
$c_0(\Lambda)$ and $c_2(\Lambda)$ algebraically \cite{sv,pbc}. The
direct use of the algebraic expressions, however, amounts to a
mere phenomenological fit.  As an application of EFT, we must
ensure that EFT counting rules are properly applied: because
our EFT Lagrangian is truncated at $p^2/\mathcal{Q}^2$, we must be
consistent with the truncation in the determination of $c_0$ and
$c_2$.  That is, $c_2(\Lambda)$ must be treated perturbatively by
neglecting the $\mathcal{O}([c_2(\Lambda)]^2)$-order
contributions.  We then obtain \cite{sv}
\begin{eqnarray}
\frac{M}{4\pi}\frac{1}{a_0} &=& \left[\frac{1}{c_0(\Lambda)} +
\frac{M}{2\pi^2}L_1\right] +
\frac{M}{\pi^2}L_3\frac{c_2(\Lambda)}{c_0(\Lambda)},
\nonumber\\
\frac{M}{16\pi} r_0 &=& \frac{c_2(\Lambda)}{c_0^2(\Lambda)} -
\frac{M}{4\pi^2}\frac{1}{\Lambda}R(0), \label{a0r0}
\end{eqnarray}
where $L_1 = \theta_1\Lambda$ and $L_3 =\theta_3\Lambda^3$.  The
numerical values of $\theta_1, \theta_3$, and $R(0)$ for large
lattices are given in Ref. \cite{sv}.  The inversion of
Eq.~(\ref{a0r0}) is, again by treating $c_2(\Lambda)$
perturbatively,
\begin{eqnarray}
c_0(\Lambda) &=&c_0^{(0)}(\Lambda)\left\{1
+\frac{r_0}{\pi}\left(\frac{M}{4\pi}\right)^2L_3\, \eta \,
[c_0^{(0)}(\Lambda)]^2\right\} \equiv c_0^{(0)}(\Lambda) + \Delta
c_0(\Lambda),
\nonumber\\
c_2(\Lambda) &=& \frac{Mr_0 }{16\pi}\, \eta \,
[c_0^{(0)}(\Lambda)]^2, \label{c02eft}
\end{eqnarray}
where $\eta = 1 + 4 R(0)/(\pi r_0 \Lambda)$, and the leading-order
$c_0(\Lambda)$, $c_0^{(0)}(\Lambda)$, is given by
\begin{equation}
c_0^{(0)}(\Lambda) =\frac{4\pi}{M}
\left(\frac{1}{a_0}-\frac{2}{\pi}L_1\right)^{-1}. \label{c00}
\end{equation}

Equations~(\ref{a0r0}) and (\ref{c02eft}) consistently include
up to $\mathcal{O}(p^2/\mathcal{Q}^2)$; their combined use
is equivalent to solving the Schr\"{o}dinger equation with the
truncated potential of Eq.~(\ref{c02eft}) {\it by treating
$c_2(\Lambda)$ perturbatively}.  That is, in this treatment, we
obtain exactly the same $a_0$ and $r_0$ as those determined
phenomenologically or obtained by solving the Schr\"{o}dinger
equation with no counting rule applied.  Because of this, the
phase shifts determined by $a_0$ and $r_0$ are also exactly the
same as those determined by the LO and NLO potentials by
consistently applying the EFT counting rule. The same EFT
treatment should also be applied to calculations of many-nucleon
systems, as we have done in this work. Note that upon the
application of the EFT counting rule, consistency is the
vital point, as is evident from the observation that $r_0$ turns
out to be negative for a certain range of $\Lambda$ if this step
is not properly applied \cite{pbc}.

For $a_0$ and $r_0$, we have used the old values of $-16.45$ fm and
$2.83$ fm, respectively \cite{Noyes1972, Breit1968}. The most recent
values are $a_0 = -18.9 \pm 0.4$ fm and $r_0 = 2.75 \pm 0.11$ fm
as quoted in Ref.~\cite{m}.  The discrepancy between the two $a_0$
values is $13 \pm 2$ \% and not negligible, but its effects
are expected to be much smaller.

As Eq.~(\ref{c00}) implies, $c_0^{(0)}$ is dominated by the
$\Lambda$ contribution because $c_0^{(0)}$ is close to the
nontrivial fixed point in the renormalization group flow
\cite{weinberg,Birse:1999}, dictated by the large magnitude of
$a_0$. Consequently, $c_0$ and $c_2$ are quite insensitive to the
exact value of $a_0$. For example, at $k_F = 60$ MeV,  using the
standard parameter set of Table I, the NLO $c_0/(a^3t)$ and
$c_2/(a^5t)$ differ by 1.8\% and 1.4\%, respectively, between the
old and the most recent values of $a_0$ and $r_0$.
The corresponding LO $c_0/(a^3t)$ differs by 1.4\% between them.

Generally some $\Lambda$ contributions must cancel in
calculating observables, so that their values are independent of
the regularization procedure. But the closeness to the fixed point
suggests the cancellation to be effectively small in this case.
Although repeating our entire calculations is quite time consuming
and unrealistic at present, we have performed a limited, test
LO calculation at $k_F = 60$ MeV for $N_s = 6^3$ and $n=1/4$. We
find $\Delta$ differs by about 2\%, in the same order of the
statistical uncertainties of the Monte Carlo calculation: $\Delta
= 0.63(1)$ and $= 0.64(3)$ MeV for $a_0=-16.45$ and $= -18.9$ fm,
respectively.  This finding also confirms the following
observation: in the accompanying paper \cite{AS-uni}, we report
the determination of various quantities at the unitary limit ($|a_0|
\rightarrow \infty$ with $r_0 = 0$) by making the extrapolation
$\eta \equiv 1/(a_0 k_F) \rightarrow 0$. By taking the $\eta$
variation to be an $a_0$ variation, we find that the above
discrepancy in $\Delta$ is 2.2\% for $k_F = 60$ MeV and decreases
as $k_F$ gets larger and increases as $k_F$ gets smaller.

\section{Physical Sizes of a Neutron Pair and Computational Lattice}

A measure of the size of an interacting neutron pair (a Cooper
pair) in the superfluid state, $\xi_{\rm cp}$, is \cite{abook}
\begin{equation}
 \xi_{\rm cp} = \frac{\hbar^2 k_F}{M \Delta}.
\end{equation}
$\xi_{\rm cp}$ must be smaller than the dimension of the cubic
lattice, as a necessary condition for the simulation of the
collective state (but clearly not a sufficient one). Table
\ref{Table:ratio_coherence_length} shows that $\xi_{\rm cp}$ is indeed
much smaller than the dimension of the lattice $a N_s^{1/3}$,
except for the marginal case of $N_s = 4^3$. Note that
$\xi_{\rm cp}$ depends on $a$ and $L$ through the $n$ dependence of
$\Delta$. The $a$ and $N_s$ dependence of $\xi_{\rm cp}$ through
$\Delta$ is weak, as seen in Sec.~IV A. In the table, we
list $\xi_{\rm cp}$ for $N_s = 4^3$ and $n = 1/4$, for simplicity.

\begin{table}[htbp]
\caption[Physical sizes of a neutron pair and computational
lattice] {Physical sizes of a neutron pair and computational
lattices.} \label{Table:ratio_coherence_length}
    \begin{center}
    \begin{tabular}{ccccccc}
      \hline
      \hline
      $k_F$ (MeV) & $\xi_{cp}$ (fm) & $a$ (fm) &
      $a N_s^{1/3}$ ($N_s=4^3$) & $a N_s^{1/3}$ ($N_s=6^3$)
      & $a N_s^{1/3}$ ($N_s=8^3$) & $a N_s^{1/3}$ ($N_s=10^3$)\\
      \hline
       \, 15 &$1.3\times10^2$ & 25.64 & 102.6 & 153.8 & 205.1 & 256.4  \\
       \, 30 &  47            & 12.82 &  \, 51.3 &  \, 76.9 & 102.6 & 128.2  \\
       \, 45 &  28            & \, 8.55 &  \, 34.2 &  \, 51.3 &  \, 68.4 &  \, 85.5  \\
       \, 60 &  21            & \, 6.41 &  \, 25.7 &  \, 38.5 &  \, 51.3 &  \, 64.1  \\
       \, 90 &  11            & \, 4.27 &  \, 17.1 &  \, 25.6 &  \, 34.2 &  \, 42.7  \\
      120 &  8.6           &  \, 3.21 &  \, 12.8 &  \, 19.2 &  \, 25.6 &  \, 32.1  \\
      \hline
      \hline
    \end{tabular}
  \end{center}
\end{table}

\section{Technical Details of Monte Carlo Computation}
\label{Preliminary_Considerations}

In this appendix, we discuss some technical details of the setup
for the implementation of our lattice calculations.

\subsection{Parameter values}

The parameter set for lattice sizes is the following: the
number of spatial lattice sites used are
\begin{equation}
  N_s = 4^3, \ 6^3, \ 8^3, \ {\rm and } \ 10^3,
\end{equation}
so as to extrapolate the data into the thermodynamic limit
($N_s \rightarrow \infty$); the number of temporal lattice sites
is
\begin{equation}
  4 \le N_t \le 128 ,
\end{equation}
where the discretization size of the temporal lattice is the same
in Ref.~\cite{Sewer} as
\begin{equation}
  \Delta N_t = \frac{0.125}{t} .
\end{equation}

The typical example of one production run is as follows.
Because the method of grand canonical ensemble is used, $\mu$ is
fixed in each run. The thermal observable for the desired density
$\rho$ is interpolated from a few sets of the observables
calculated at different $\mu$. About $1000$-$10000$ samples are
accumulated to obtain statistics with a precision of several
percent.

\subsection{Determinantal quantum Monte Carlo}
\label{sec:Setup_for_DQMC}

\subsubsection{Temporal lattice spacing}

\begin{figure}[htbp]
\begin{center}
\includegraphics[width=100mm]{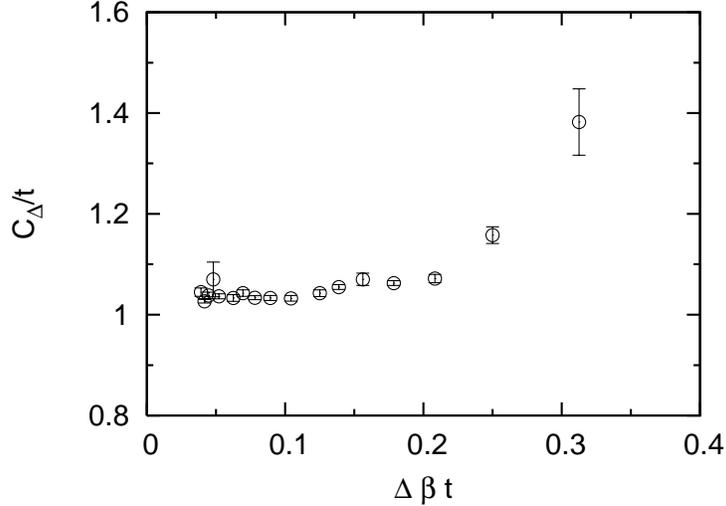}
\caption{Pair correlation function $C_\Delta$ as a function of
temporal lattice spacing $\Delta \beta$, with $N_s = 4^3$ at $k_F
= 30$ MeV in the unit of hopping amplitude $t$ in the DQMC
calculation.} \label{Fig:beta_CDelta.DQMC}
\end{center}
\end{figure}

To choose $\Delta \beta$, we need to know how the
expectation values of thermal observables are affected by the
choice. Figure~\ref{Fig:beta_CDelta.DQMC} illustrates the
dependence of $\Delta \beta$ on the thermal observable $C_\Delta$
in our DQMC calculation. The data have been taken with $\mu/t = 0$
and $N_s=4^3$ at $k_F = 30$ MeV. The figure is a typical example,
and we have observed similar results with other thermal
observables and parameter values.

From Fig.~\ref{Fig:beta_CDelta.DQMC}, we see that the expectation
values of thermal observables are affected only a little for
$\Delta \beta \lsim 0.2 t$, confirming that the choice employed in
the previous DQMC calculation similar to ours \cite{dosSantos2003}
is indeed reasonable, and so we adopted this choice.

\subsubsection{Prethermalization steps}

At the start of sampling, we generate the initial configuration of
the auxiliary fields $\chi$. In our DQMC calculation, we use the
hot start, in which a random (disordered) configuration is used,
instead of the cold start using a uniform (ordered) configuration.
Following the start, we must take a sufficient number of
prethermalization steps to obtain the equilibrium
configurations, statistically independent from the initial
configuration in the Markov chain.

\begin{figure}[htbp]
\begin{center}
\includegraphics[width=100mm]{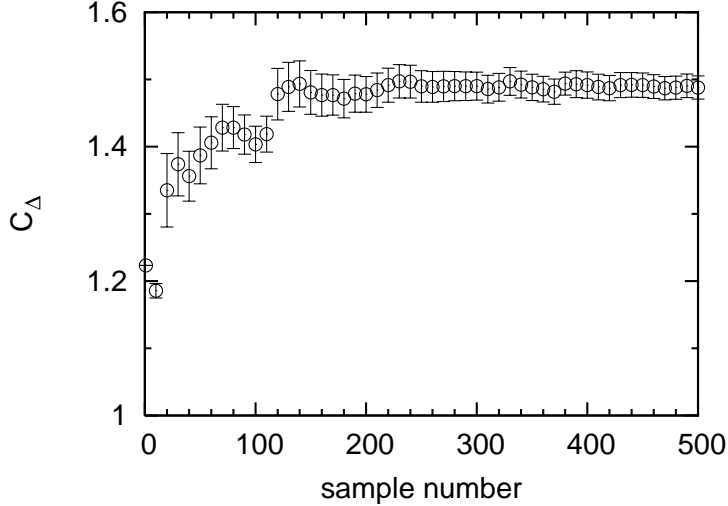}
\caption{The pair correlation function $C_\Delta$ as a function of
the sample number at $N_s = 4^3$ and $N_t = 12$ in the unit of
hopping amplitude $t$ in our DQMC calculation.}
\label{Fig:pretherm_CDelta.DQMC}
\end{center}
\end{figure}

Figure~\ref{Fig:pretherm_CDelta.DQMC} illustrates the dependence
of the sample number on the thermal observable $C_\Delta$ in our
DQMC calculation. The data have been taken with $\mu/t = -1.83$ at
$N_s=4^3$ and $N_t=12$.  The figure shows that the equilibrium
starts to be reached after $100$-$150$ samples.  Similar results are
observed with other observables and for other parameter values.

\subsubsection{Thermalization steps and autocorrelations}

To ensure statistically independent configurations, we
must take thermalization (decorrelation) steps between sample
takings. We determine the number of the thermalization steps by
monitoring the autocorrelation.  The autocorrelation for $k$
conservative samples of the observable $O$, $C_{O}(k)$, is of the
standard form
\begin{equation}
  C_{O}(k) = \frac{\langle O_i O_{i+k}\rangle - \langle O_i \rangle^2}
  {\langle O_i^2 \rangle - \langle O_i \rangle^2}
\end{equation}
where $\langle \cdots \rangle$ denotes the average over the random
walk labeled with $i$, for example,
\begin{equation}
  \langle O_i O_{i+k} \rangle
  \equiv \frac{1}{N-1}\sum_{i=1}^{N-k} O(X_i) O(X_{i+k}).
\end{equation}
The condition of no correlation is $C_O \sim 0$, but in practical
terms $C_O \lsim 0.1$ is recommended \cite{Koonin:1990}, and thus
we ensure $C_O$ to be less than 10\% .

\begin{figure}[htbp]
\begin{center}
\includegraphics[width=100mm]{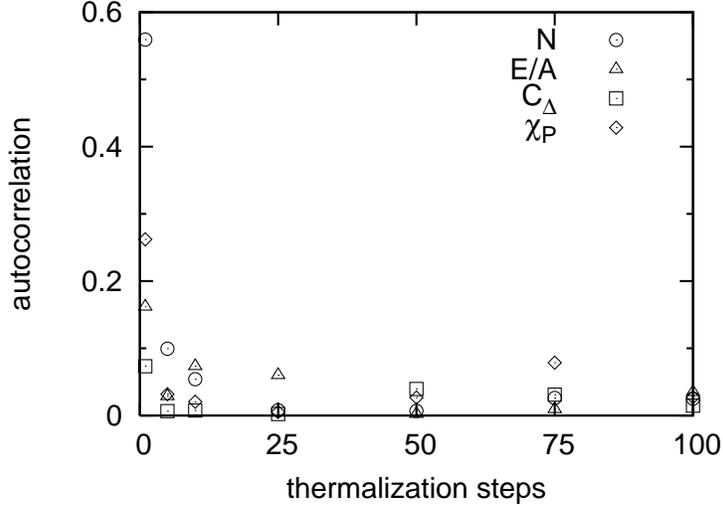}
\caption{Autocorrelation as a function of thermalization
steps between samples taken with the number of spatial lattice
sites $N_s = 4^3$ and of temporal lattice sites $N_t = 12$ at the
Fermi momentum $k_F = 30$ MeV in our DQMC calculation.}
\label{Fig:autocorrelation.DQMC}
\end{center}
\end{figure}

A typical case of the autocorrelations for some observables is
shown in Fig.~\ref{Fig:autocorrelation.DQMC} with the parameter
set ($N_s = 4^3$, $N_t = 12$, and $k_F = 30$ MeV).  The
autocorrelations are seen to be less than 0.1 for more than ten
thermalization steps between samples.

\subsection{Systematic error of the DQMC}

Here, we discuss the systematic uncertainties of the DQMC besides
the statistical ones due to data sampling. After ensuring the
independence between samples by keeping the autocorrelations of
thermal observables small enough as described in Appendix A 2, the
systematic error of the DQMC on observables solely comes
from the size of the discretization of the time slice $\Delta
\beta$, which is related to the inverse of temperature $\beta
\equiv N_t \Delta \beta$.

For confirming the consistency of our DQMC calculation with
others, we compare $T_c/t$ with that in Refs.~\cite{Sewer,
Sewer:PhDthesis} over the various interaction strengths
$c_0/(a^3t)$ at fixed temporal lattice spacing $\Delta \beta =
0.125/t$, which has been commonly used in the condensed-matter
physics. For estimating the systematic errors caused by finite
$\Delta \beta$, the $\Delta \beta$ dependence of $\Delta$,
$T_c$, and $T^\ast$ have also been further examined.

By these preliminary DQMC calculations, we can ensure the
consistencies of DQMC calculations with those in other literature.
The systematic uncertainties caused by our calculations with
finite $\Delta \beta$ amount to around 10\%.

\subsubsection{Comparison of $T_c (c_0/(a^3t))$ with other work}

First we ensure that our DQMC calculation at finite $\Delta
\beta$ is consistent with other literature. Figure
\ref{Fig:BCS_BEC_crossover} is the critical temperature $T_c$ as a
function of interaction strength $|c_0|/(a^3t)$ at the
quarter-filling ($n = 1/2$) in $N_s = 6^3$. $T_c$ is obtained
through the inflexion point of the curve of pair correlation
function $C_\Delta$. The parameters used in the calculations
are $\Delta \beta = 0.125/t$, $N_{\rm pretherm} = 200$, $N_{\rm therm} =
50$, $N_{\rm sample} = 1000$-$2000$. Our $T_c(|c_0|/(a^3t))$ over
the interaction strength ranging between BCS and BEC limits is in
good agreement with Refs.~\cite{Sewer, Sewer:PhDthesis} of the same
setup within around 5\% of errors, which is within the DQMC
results in other literature, ranging around 10\% at half-filling
($n=1$) as shown in the left panel of Fig.~5.13 in
Ref.~\cite{Sewer:PhDthesis}.

\subsubsection{Dependence of thermal observables on $\Delta \beta$}

\begin{figure}[htbp]
\begin{center}
\includegraphics[width=100mm]{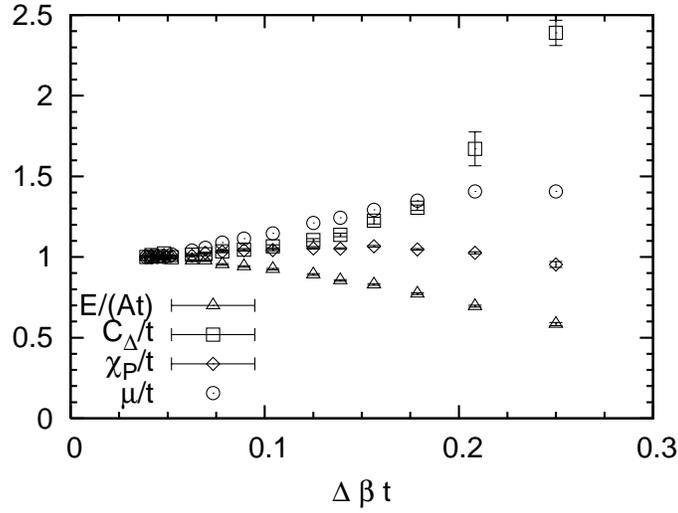}
\caption{$\Delta \beta$ dependence of the ratio of energy per
particle $E/A$, pair correlation function $C_\Delta$, 
Pauli spin susceptibility $\chi_P$, and
chemical potential $\mu$ to those at $\Delta \beta
\rightarrow 0$ at $T/t = 0.4$ with the interaction strength
$c_0/(a^3t) = -6.0$ at the one-eighth filling ($n = 1/4$) in the
dimensionless unit.} \label{Fig:beta_obs}
\end{center}
\end{figure}

Now that our DQMC calculations with finite $\Delta \beta$ are
confirmed within around 5\% of the differences, we have to
consider the systematic error from the discretization of temporal
direction $\Delta \beta$. Figure \ref{Fig:beta_obs} shows the
dependence of various thermal observables on $\Delta \beta$ by
fixing $T/t = 1/(N_t \Delta \beta t) = 0.4$. The expectation
values of thermal observables are obtained by $1000$-$2000$
samples with $N_{\rm pretherm} = 200$ and $N_{\rm therm} = 100$ at
the one-eighth filling $(n = 1/4)$. In Fig.~\ref{Fig:beta_obs},
we take the ratio of thermal observables at $\Delta \beta =
0.125/t$ to those at the continuum limit of the temporal direction
$\Delta \beta \rightarrow 0$ to make the deviations easily
visible. As summarized in Table \ref{table:ratio_obs}, the
differences of the observables with $\Delta \tau = 0.125/t$ and
$\Delta \beta \rightarrow 0$ are around 5\% (for $\chi_P$),
10\% (for $C_\Delta$ and $E/A$), and 20\% (for $\mu$). Note that
we use only $C_\Delta$ and $\chi_P$ for obtaining $T_c$ and
$T^\ast$ in this work.

\begin{table}[htbp]
\caption[Ratio of thermal observables at $\Delta \beta = 0.125/t$
to $\Delta \beta \rightarrow 0$.] {Ratio of thermal observables.}
\label{table:ratio_obs}
    \begin{center}
    \begin{tabular}{cccc}
      \hline
      \hline
      ${\cal O}$ & ${\cal O}(\Delta \beta = 0.125/t)$ &
      ${\cal O}(\Delta \beta \rightarrow 0)$ &
      ${\cal O}(\Delta \beta = 0.125/t)/{\cal O}(\Delta \beta \rightarrow 0)$        \\
      \hline
      $E/(At)$     & \, 1.449(8) & \, 1.625(8)  & \, 0.892(9) \\
      $C_\Delta$   & 1.45(1)   & 1.32(1)   & 1.10(2) \\
      $\chi_P$     & \, 0.207(9)  & \, \, 0.1965(9) & 1.05(5) \\
      $\mu/t$      & 1.49(1)    & 1.23(1)    & 1.21(1)  \\
      \hline
      \hline
    \end{tabular}
  \end{center}
\end{table}

\begin{figure}[htbp]
\begin{center}
\includegraphics[width=100mm]{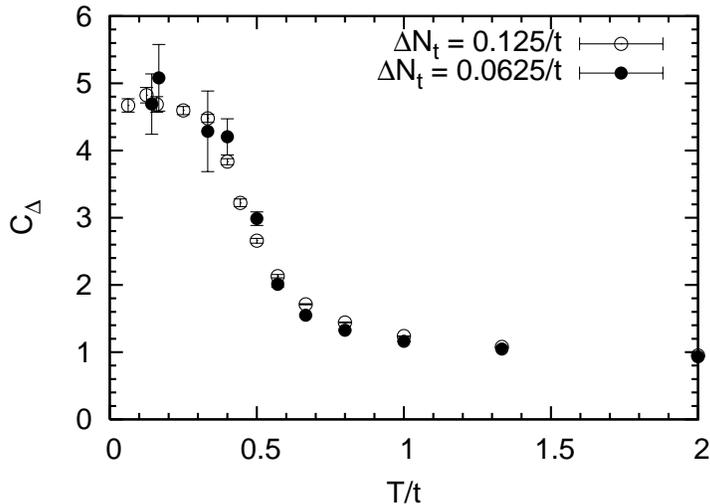}
\caption{Pair correlation function $C_\Delta$ as a function of
temperature $T$ in the unit of hopping amplitude $t$ at different
$\Delta \beta$ at $k_F = 30$ MeV, $N_s = 4^3$, and $n = 1/4$. The
open and solid circles with statistical errors are the
results at LO and NLO, respectively. } \label{Fig:CDelta_DeltaN}
\end{center}
\end{figure}

\begin{figure}[htbp]
\begin{center}
\includegraphics[width=100mm]{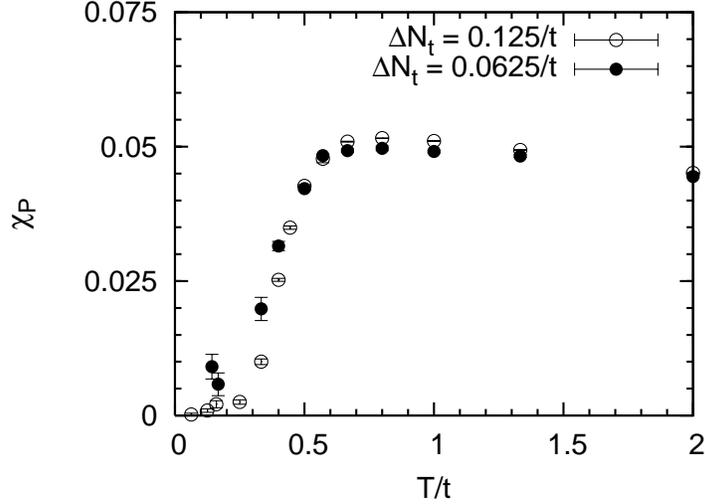}
\caption{Pauli spin susceptibility $\chi_P$ as a function of
temperature $T$ in the unit of hopping amplitude $t$ at different
$\Delta \beta$ at $k_F = 30$ MeV, $N_s = 4^3$, and $n = 1/4$. The
open and solid circles with statistical errors are the
results at LO and NLO, respectively.} \label{Fig:chiP_DeltaN}
\end{center}
\end{figure}

Next we examine the influence of finite $\Delta \beta$ on $T_c$
and $T^\ast$. Figures \ref{Fig:CDelta_DeltaN} and
\ref{Fig:chiP_DeltaN} summarize the effect of the finite $\Delta
\beta$ on $T_c$ and $T^\ast$. As seen in those figures, $T_c/t =
0.45(1)$ MeV and $T^\ast/t = 0.87(2)$ MeV for $\Delta \beta =
0.125/t$, and $T_c/t = 0.47(2)$ MeV and $T^\ast/t = 0.79(2)$ MeV
for $\Delta \beta = 0.0625/t$. The quantities in the parentheses
indicate the statistical uncertainties. The deviations in $T_c$
and $T^\ast$ without the statistical errors are around 5\% and
10\%, respectively. We have to count on these discrepancies
of around 10\% as the systematic error of our final results
besides the statistical error.

As described in Sec.~III, we use $P_s$ for an estimation of
$\Delta$. The constant tails of $P_s$ at the large separation
of pairs are $P_s(\Delta \beta = 0.125/t) = 0.02784(46)$ and
$P_s(\Delta \beta = 0.0625/t) = 0.0295(24)$ at $k_F = 30$ MeV. The
resultant pairing gaps extracted from $P_s$ through $\Delta = c_0
\sqrt{P_s}$ with $c_0 = 0.8012$ MeV are $\Delta (\Delta \beta =
0.125/t) = 0.1337(11)$ MeV and $\Delta (\Delta \beta = 0.0625/t) =
0.1377(56)$ MeV. The deviation between them without the
statistical errors quoted by the parentheses is 0.004 MeV, which
results in around 3\% of the systematic error.


\end{document}